\documentclass[11pt]{article}
\usepackage{comment}
\usepackage[margin=2.5cm]{geometry}
\usepackage{amsmath,amssymb,extarrows,mathtools,graphicx,subfigure,setspace,bbold,physics}
\usepackage{cite}
\usepackage{slashed}
\usepackage[dvipsnames]{xcolor}
\usepackage{hyperref}
\hypersetup{colorlinks=true, linkcolor=blue, citecolor=magenta, linktoc=page}
\usepackage{tikz}
\usepackage{color}
\usepackage{arydshln,leftidx,mathtools}
\usetikzlibrary{decorations.pathreplacing}
\numberwithin{equation}{section}
\newcommand{\be}{\begin{equation}}
	\newcommand{\bea}{\begin{eqnarray}}
		\newcommand{\eea}{\end{eqnarray}}
	\newcommand{\ba}{\begin{align}}
		\newcommand{\ea}{\end{align}}
	\newcommand{\ee}{\end{equation}}

\usepackage{chngcntr}
\begin{document}
	
\begin{titlepage}
	\thispagestyle{empty}

	\begin{flushright}
		IPM/P.A-745
	\end{flushright}
	
	\vspace{.4cm}
	\begin{center}
		\noindent{\Large \textbf{Holographic Study of Reflected Entropy in Anisotropic Theories}}\\
		
		\vspace*{15mm}
		\vspace*{1mm}
		\vspace*{1mm}
		{Mohammad Javad Vasli${}^{\ast}$, M. Reza Mohammadi Mozaffar${}^{\ast}$ \\ Komeil Babaei Velni${}^{\ast}$and  Mohammad Sahraei${}^{\dagger,*}$}
		
		\vspace*{1cm}
		
		{\it  ${}^\ast$ Department of Physics, University of Guilan,
			P.O. Box 41335-1914, Rasht, Iran\\
			${}^\dagger$ School of Particles and Accelerators,
			Institute for Research in Fundamental Sciences (IPM),
			P.O. Box 19395-5531, Tehran, Iran
		}
		
		\vspace*{0.5cm}
		{E-mails: {\tt vasli@phd.guilan.ac.ir, mmohammadi@guilan.ac.ir, babaeivelni@guilan.ac.ir, msahraei@ipm.ir}}
		
		\vspace*{1cm}
%		\maketitle
	\end{center}
	
	\begin{abstract}

We evaluate reflected entropy in certain anisotropic boundary theories dual to nonrelativistic geometries using holography. It is proposed that this quantity is proportional to the minimal area of the
entanglement wedge cross section. Using this prescription, we study in detail the effect
of anisotropy on reflected entropy and other holographic entanglement measures. In
particular, we study the discontinuous phase transition of this quantity for a symmetric configuration consisting of two disjoint strips. We find that in the specific regimes of the parameter space
the critical separation is an increasing function of the anisotropy parameter and hence the
correlation between the subregions becomes more pronounced. We carefully examine how these results are consistent with the behavior of other correlation measures including the mutual information. Finally, we show that the structure of the universal terms of entanglement entropy is corrected depending on the orientation of the entangling region with respect to the anisotropic direction.

\end{abstract}
	
\end{titlepage}
\newpage

\tableofcontents
\noindent
\hrulefill

\onehalfspacing
%%%%%%%%%%%%%%%%%%%%%%%%%%

\section{Introduction}\label{intro}
In recent years, the holographic framework allows us to quantitatively study the 
fascinating connections between quantum information and quantum gravity. In this context, different quantum information measures and their holographic counterparts have proved very useful for developing our understanding of the gauge/gravity correspondence, e.g., entanglement entropy and computational complexity\cite{Ryu:2006bv,Brown:2015bva}. In particular, the Ryu-Takayanagi (RT) prescription has proven immensely useful for investigating this connection in a robust manner, by constructing a geometrical realization of the entanglement entropy (EE) for a spatial subregion in the boundary field theory. Let us recall that EE has emerged as an interesting theoretical quantity which provides new insights into a variety of topics in physics ranging from quantum information theory to high energy physics(see \cite{Casini:2009sr,Nishioka:2018khk} for reviews). Moreover, the entanglement entropy is the unique quantity which measures the amount of quantum entanglement between two subsystems for a given pure state. In this case, assuming that the total Hilbert space takes a direct product form of two Hilbert spaces of the subsystems, \textit{i.e.}, $\mathcal{H}=\mathcal{H}_A\otimes \mathcal{H}_{\bar{A}}$, the corresponding EE of the subsystem $A$ is given as follows
\begin{eqnarray}\label{EE}
S_A=-{\rm Tr}_{A} \;\rho_A\log\rho_A,
\end{eqnarray}
where $\rho_A$ is the reduced density matrix defined as $\rho_A={\rm Tr}_{\bar{A}}|\psi\rangle\langle\psi|$ and $|\psi\rangle$ denotes the corresponding pure state. The holographic counterpart of eq. \eqref{EE} can be obtained using RT prescription which states that EE is dual to the area of a minimal codimension-two bulk hypersurface $\Gamma_A$ which is homologous to the boundary region $A$, \textit{i.e.}, \cite{Ryu:2006bv}
\begin{eqnarray}\label{HEE}
S_A=\frac{{\rm min}\left({\rm area} \;\Gamma_A\right)}{4G_N}.
\end{eqnarray}
Hence, in strongly coupled quantum field theories with holographic duals, computing EE reduces to a geometric problem of finding minimal hypersurfaces satisfying suitable boundary conditions. This proposal has stimulated a wide variety of research efforts investigating the properties and applications of holographic entanglement entropy (HEE), \textit{e.g.}, see\cite{Rangamani:2016dms,Nishioka:2009un} for reviews.

Further, EE fails to be a good measure of quantum entanglement or correlations between the subsystems for mixed states.  
%Despite being relatively well understood for pure states, only recently entanglement measures have been developed for such classes of states. 
A variety of correlation measures for such classes of states have been developed, \textit{e.g.}, logarithmic negativity\cite{Plenio:2005cwa}, entanglement of purification \cite{Terhal:2002}, odd entropy \cite{Tamaoka:2018ned} and reflected entropy \cite{Dutta:2019gen}. Much of our
analysis in this paper will focus on studying reflected entropy in specific holographic settings, so we proceed
by reviewing its definition. Consider a mixed state $\rho=\sum_i p_i|\rho_i\rangle\langle\rho_i|$ in $\mathcal{H}_A\otimes \mathcal{H}_B$. The canonical purification is defined on a doubled Hilbert space $\mathcal{H}_A\otimes \mathcal{H}_{A'}\otimes\mathcal{H}_B\otimes\mathcal{H}_{B'}$ and is given by 
\begin{eqnarray}\label{canonical}
|\sqrt{\rho}\rangle=\sum_i \sqrt{p_i}|\rho_i\rangle\otimes|\rho_i\rangle.
\end{eqnarray}
Now the reflected entropy is the corresponding EE of the subsystem $AA'$, \textit{i.e.}, 
\begin{eqnarray}\label{RE}
S_R(A, B)=-{\rm Tr}\rho_{AA'} \log\rho_{AA'},
\end{eqnarray}
where $\rho_{AA'}={\rm Tr}_{BB'}|\sqrt{\rho}\rangle\langle\sqrt{\rho}|$. Clearly, the above definition reduces to EE when $\rho$ is pure. There are several interesting inequalities which the reflected entropy satisfies generally, \textit{e.g.},
\bea\label{iequalities}
&I(A, B)\leq S_R(A, B)\leq 2\;{\rm min}\{S_A, S_B\},\nonumber\\
&I(A, B)+I(A, C)\leq S_R(A, B\cup C),
\eea
where $I(A, B)$ is the mutual information between $A$ and $B$ given as follows
\bea\label{HMI}
I(A, B)=S_A+S_B-S_{A\cup B}.
\eea

In \cite{Dutta:2019gen} the authors provided an interesting holographic interpretation of the canonical purification and also proposed a dual counterpart for the reflected entropy which is the minimal cross sectional area of the entanglement wedge. Before we proceed further, let us recall that the entanglement wedge is the bulk region corresponding
to the reduced density matrix $\rho_A$ and whose boundary is $A\cup \Gamma_A$. Considering a spatial boundary region consists of two disjoint parts $A$ and $B$ and denoting the cross sectional area of the entanglement wedge by $\Sigma_{A\cup B}$ the corresponding reflected entropy is given by
\begin{eqnarray}\label{HRE}
S_R(A, B)=\frac{{\rm min}\left({\rm area} \;\Sigma_{A\cup B}\right)}{2G_N}.
\end{eqnarray}
A key feature of the above proposal is that the holographic reflected entropy presents a discontinuous phase transition from zero to positive values as the two subregions get closer. This transition is due to the competition between a connected and a disconnected configuration for the entanglement wedge. Indeed, for large separations where the disconnected configuration is favored, $\Sigma_{A\cup B}$ becomes empty and the corresponding reflected entropy vanishes. Let us recall that there exist other measures which seem to be dual to entanglement wedge cross section (EWCS). These proposals can be summarized as follows\cite{Tamaoka:2018ned,Dutta:2019gen,Kusuki:2019zsp}
\begin{eqnarray}\label{ewsr}
E_W(A, B)=\frac{S_R(A, B)}{2}=\frac{\mathcal{E}(A, B)}{\chi_d}=S_O(A, B)-S(A\cup B),
\end{eqnarray}
where $S_R$, $\mathcal{E}$ and $S_O$ are reflected entropy, logarithmic negativity and odd entropy respectively.\footnote{Let us also mention that the connection between $E_W$ and $\mathcal{E}$ was recently called into question by \cite{Dong:2021clv}.} Here $\chi_d$ is a constant which depends on the dimension of the spacetime. These proposals has since been the subject of a large body of work \cite{Takayanagi:2017knl,Nguyen:2017yqw,Kusuki:2019zsp,Hirai:2018jwy,BabaeiVelni:2019pkw,Jokela:2019ebz,Umemoto:2019jlz,Akers:2019gcv,Amrahi:2020jqg,Chakrabortty:2020ptb,Saha:2021kwq,1810.00420,1907.06646,Jeong:2019xdr,2001.05501,BabaeiVelni:2020wfl,Moosa:2020vcs,Boruch:2020wbe,Basu:2022nds,Basak:2022cjs}. Further, a wide variety of recent research efforts investigating the properties of the corresponding measures from the perspective of the boundary theory have also appeared in \cite{1909.06790,Mollabashi:2020ifv,Berthiere:2020ihq,Bueno:2020fle,Camargo:2021aiq,Wen:2021qgx,Hayden:2021gno,Akers:2021pvd,Bueno:2020vnx,Camargo:2022mme}.

Our goal in this paper is to present another step in this research program, in which we investigate the behavior of reflected entropy in anisotropic systems with strong interactions by means of holography. Let us recall that anisotropic holographic models have already been extensively studied in the context of AdS/QCD to scan the QCD phase diagram and also to investigate different aspects of quark-gluon plasma which is produced in relativistic heavy ion collisions, \textit{e.g.}, \cite{Mateos:2011tv,Rougemont:2015oea,Rebhan:2011vd,Mateos:2011ix,Giataganas:2012zy,Chernicoff:2012iq,Ali-Akbari:2013txa}. On the other hand, in the context of AdS/CMT, anisotropic holographic models appear in many examples of quantum criticality in condensed matter physics with non-relativistic fixed points \cite{Pal:2009yp,Azeyanagi:2009pr}. Further, some investigations attempting to better understand the behavior of different holographic entanglement measures in anisotropic backgrounds have also appeared in \cite{Narayan:2012ks,Jahnke:2017iwi,Mahapatra:2019uql,Arefeva:2020uec,Jain:2020rbb,Sahraei:2021wqn}. In this paper, we aim to 
provide a detailed study of the influence of anisotropy on the behavior of reflected entropy. An especially interesting question concerns how the phase transition of this quantity is affected by anisotropy. We will also discuss how our results are comparable with the behavior of other correlation measures including the holographic mutual information (HMI).

% In particular, we are interested in various ...... For this purpose,

The remainder of our paper is organized as follows: In section \ref{sec:setup}, we give the general framework
in which we are working, establishing our notation and the general form of the HEE and reflected entropy
functionals in a static anisotropic background. In section \ref{sec:condecon}, we consider an anisotropic geometry whose dual state exhibits confinement-deconfinement phase transition and study the properties of reflected entropy numerically. To get a better understanding of the results, we will also compare the behavior of reflected entropy to other correlation measures including HEE and HMI. In section \ref{sec:EAD}, we extend our studies to a family of axion-dilaton gravity theories underlying solution breaks isotropy while preserving
translation invariance. By tuning the dilaton potential, we study the influence of anisotropy on reflected entropy in different backgrounds. In the latter case we present a combination of numerical and analytic results on the scaling of different correlation measures. Next, we study a specific geometry with anisotropic Lifshitz scale invariance in section \ref{sec:Lifshitz}. We review our main results and discuss their physical implications in section \ref{sec:diss}, where we also indicate some future directions.

\section{Set-up}\label{sec:setup}
\begin{figure}[h]
\begin{center}
\includegraphics[scale=0.7]{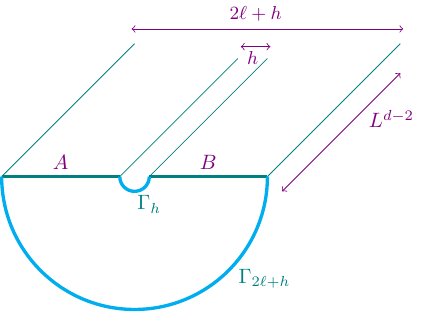}
\hspace*{1cm}
\includegraphics[scale=0.9]{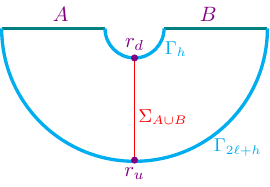}
\end{center}
\caption{\textit{Left}: Schematic minimal hypersurfaces for computing $S_{A\cup B}$ in connected configuration. \textit{Right}: The minimal cross section of the entanglement wedge, $\Sigma$ in red. Here we only show the connected configuration where the reflected entropy is non-zero.}
\label{fig:regions}
\end{figure}

In this section, we briefly review some preliminaries to construct the holographic reflected entropy functional in generic anisotropic geometries. We focus our analysis on the special case of a five-dimensional bulk geometry because the interesting qualitative features of the reflected entropy are independent of the dimensionality of the boundary field theory. In this case, the general form of an anisotropic background can be written as\footnote{Note that using the reparametrization
invariance one can fix $G_1(r)$ and $G_2(r)$ in eq. \eqref{genmetric} and once this is done, $b(r)$ cannot be set to unity in general. In section \ref{secplasma} we consider an specific background with $b(r)\neq 1$.}  
\begin{equation}\label{genmetric}
ds^2=\frac{R^2}{r^2}H(r) \left(  -f(r)b(r)dt^2+ \sum_{n=1}^{3}G_n(r) dx_n^2+\frac{dr^2}{f(r)}\right),
\end{equation}
%\begin{equation}\label{genmetric}
%ds^2=\frac{R^2}{r^2}H(r) \sum_{n=0}^{4}G_n(r) dx_n^2.
%\end{equation}
where $R$ is the curvature radius. Without loss of generality we will from now on consider $R=1$. In order to investigate the effect of anisotropy on the reflected entropy we consider the simplest boundary entangling region consisting of two disjoint long narrow strips with equal width $\ell$ separated by $h$ on a constant time slice (see figure \ref{fig:regions}). 
Further, to examine the effects of changing the direction of the strip, we lay entangling region in some arbitrary direction using rotation with Euler angles as follows
\begin{align}
x_i(\xi)&=\sum_{j=1,2,3} a_{ij} (\alpha,\beta,\gamma ) \xi_{j},\qquad i=1,2,3
\end{align}
where $a_{ij}$ is the entry of the rotation matrix and $\alpha, \beta$ and $\gamma$ denote the angles of rotation around $x, y$ and $z$ directions, respectively. For simplicity, we will only consider rotations around the $y$-axis. Considering the width of the strip along the $\xi_1$ direction and using eq. \eqref{genmetric}, the entropy functional is then given by the following expression
\begin{equation}\label{S2}
S=\frac{L^2}{4G_N } \int \frac{H^{\frac32}(r)}{r^3} \sqrt{ \mathcal{G}(r) \xi_1'^2 +\frac{ \mathcal{T}(r,\beta)}{f(r)}  } d r,
\end{equation}
where the prime indicates derivative with respect to $r$ and we have defined
\begin{equation}
\mathcal{T}(r,\beta) =G_2 (r)\left(G_1 (r) \sin^2 \beta +G_3(r)  \cos^2 \beta\right),\hspace*{1cm}\mathcal{G}(r)= G_1(r) G_2(r) G_3(r).
\end{equation}
Further, using the equation of motion, the width of the entangling region and HEE can be written as follows
\begin{equation}\label{Gl}
\ell =2 \int_{0}^{r_t} \frac{\sqrt{\mathcal{T}(r,\beta) }}{\sqrt{f(r)\mathcal{G}(r)}  \sqrt{\frac{r_t^6}{r^6}\frac{ \mathcal{G}(r) H(r)^3}{\mathcal{G}(r_t) H({r_t})^3}-1}} dr.
\end{equation}
\begin{equation}\label{GS}
S=\frac{L^2}{2 G_N } \int_{\epsilon}^{r_t} \frac{{r_t}^3\sqrt{\mathcal{G}(r)} H(r)^3}{r^3 \sqrt{f(r)}} \sqrt{\frac{\mathcal{T}(r,\beta)}{{r_t}^6 \mathcal{G}(r)H(r)^3-r^6 \mathcal{G}(r_t) H({r_t})^3}} dr,
\end{equation}
where $r_t$ denotes the turning point of the minimal hypersurface and we regulate the calculation of the entropy in the standard way by introducing a cutoff surface at $r=\epsilon$.

Let us now turn to the computation of the reflected entropy in this setup using eq. \eqref{HRE}. Due to the symmetry of the configuration that we have chosen, $\Sigma_{A\cup B}$, lies entirely on $\xi_1 = 0$ slice and as a consequence, from eq. \eqref{genmetric}, we find the reflected entropy to be
\begin{equation}\label{GEW}
S_R=\frac{L^2}{2 G_N } \int_{r_d}^{r_u} \frac{H^{\frac32}(r)}{r^3} \sqrt{ \frac{ \mathcal{T}(r,\beta)}{f(r)}  } dr,
\end{equation}
where $r_d$ and $r_u$ denote the corresponding turning points of $\Gamma_h$ and $\Gamma_{2\ell+h}$ respectively (see figure \ref{fig:regions}).

\section{Anisotropic Theories with Confinement-deconfinement Phase Transition} \label{sec:condecon}
The first model we consider is that of an anisotropic background which exhibits confinement-deconfinement phase transition. The corresponding metric is given as follows\cite{Aref2018}
\begin{equation}
ds^2=\frac{e^{-\frac{r^2}{2}} }{r^2} \left(-f(r)dt^2+\frac{dr^2}{f(r)}+dx^2 +r^{2-\frac{2}{\nu}} (dy^2+dz^2)\right).
\end{equation}
The explicit forms of $f(r)$ is tedious and hence we do not explicitly show the corresponding expressions here. Clearly, the strength of the anisotropy between boundary spatial directions is parametrized by $\nu$ and for $\nu=1$ we have a isotropic background. This metric is a solution to Einstein gravity coupled to a dilaton and two Maxwell fields with a nontrivial scalar potential. 
In comparing the above expression with metric \eqref{genmetric}, we should identify
\begin{equation}
G_1(r)=1,\qquad\qquad G_2(r)=G_3(r)=r^{2-\frac{2}{\nu}}, \qquad\qquad H(r)= e^{-\frac{r^2}{2}} .
\end{equation}
The corresponding expression for HEE and reflected entropy can be obtained using eqs. \eqref{GS} and \eqref{GEW} and the above identifications. Different aspects of HEE in the dual boundary theory has been studied in \cite{Arefeva:2020uec}. Before we proceed further, let us comment on a characteristic property of this geometry. Indeed, as demonstrated in \cite{Arefeva:2020uec}, for some values of $r_h$ we can find several HEE for same value of $\ell$, and based on \eqref{HEE}, We have to choose the smallest ones. The transition between different hypersurfaces shows itself as a phase transition on HEE which gives a crossover transition between confinement-deconfinement phases in the dual gauge theory.  
 Further, the thermodynamical properties of this gravitational background were studied in \cite{Aref2018} and it was shown that it has a Van der Waals-like phase transition between small and large black holes for a specific range of the boundary chemical potential. More explicitly, the thermal entropy function is multivalued for $0<\mu<\mu_{\rm crit.}(\nu)$ and becomes one-to-one for $\mu_{\rm crit.}(\nu)<\mu$.   In the small black hole phase ($0<\mu<\mu_{\rm crit.}(\nu)$), HEE becomes one-to-one and phase transition doesn't occur but in the large black hole phase ($\mu>\mu_{\rm crit.}(\nu)$), HEE is multivalued and has phase transition.
 In the next subsections we will compute the holographic entanglement measures numerically and treat these cases separately.

\subsection{Reflected Entropy for small black  hole}
In this case we choose $\mu$ such that we have Van der Waals-like phase transition. We also set $r_h=1$ throughout this section and hence we are in small black hole phase. This choice corresponds to a boundary theory with confined degrees of freedom. Also for simplicity, we have rescaled the holographic measures, \textit{i.e.}, $\{S, I\}\rightarrow \frac{4G_N}{L^{2}}\{S, I\} $ and $S_R \rightarrow \frac{2G_N}{L^{2}}S_R $.

\begin{figure}[h]
\begin{center}
\includegraphics[scale=0.45]{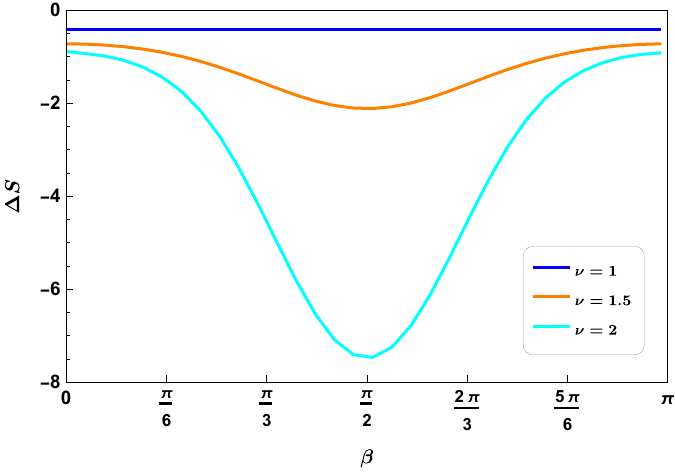}
\hspace*{0.01cm}
\includegraphics[scale=0.45]{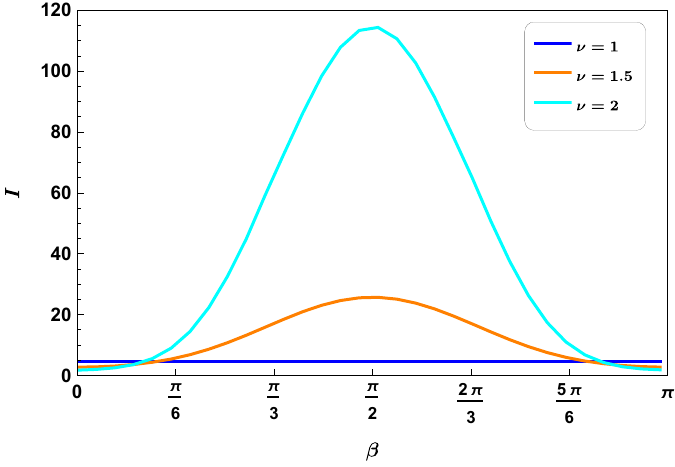}
\hspace*{0.01cm}
\includegraphics[scale=0.45]{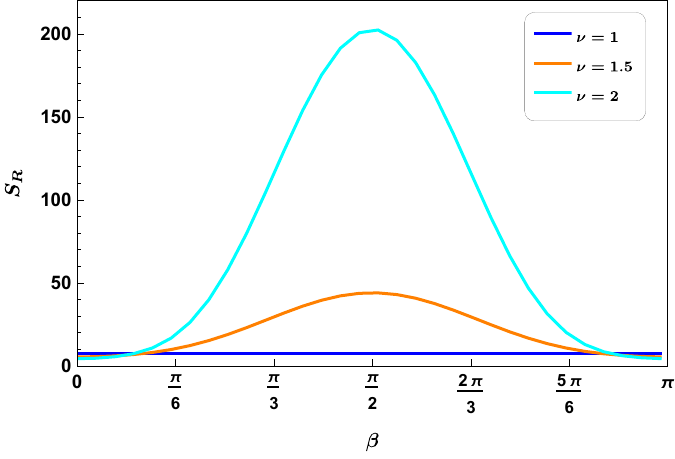}
\end{center}
\caption{Finite part of the HEE (left), HMI (middle) and reflected entropy (right) as a function of rotation angle for different values of $\nu$ with $\ell=0.4$ and $h=0.2$. The dashed straight line corresponds to isotropic case with $\nu=1$ where the measures are independent of $\beta$.}
\label{fig:numfig0}
\end{figure}

In figure \ref{fig:numfig0} we show the dependence of the HEE, HMI and reflected entropy for specific values of $\nu$ as a function of the rotation angle with $\ell=0.4$ and $h=0.2$. The dashed curve corresponds to isotropic case with $\nu=1$ where the measures are independent of the rotation angle. The left panel demonstrates the dependence of the finite part of the HEE defined as $\Delta S\equiv S-S_{\rm dis}$ on $\beta$. Here $S_{\rm dis}$ is the area of two disconnected straight lines extending from the endpoints of the boundary line segment to horizon of  black hole. Note that based on this definition the disconnected piece depends on the temperature. The corresponding area functional can be obtained by setting $\xi_1'=0$ in eq. \eqref{S2}
\begin{equation}\label{S2}
	S_{\rm dis}=\frac{L^2}{2G_N } \int_{0}^{r_h} \frac{H^{\frac32}(r)}{r^3} \sqrt{ \frac{ \mathcal{T}(r,\beta)}{f(r)}  } d r.
\end{equation}

We see that $I$ and $S_R$ have a maximum at $\beta=\pi/2$ where the HEE develops a minimum. This minimum becomes deeper and sharper for larger values of $\nu$. 
%Due to the periodic behavior of these measures, for $\pi/2\leq \beta \leq \pi$ the

\begin{figure}[h]
\begin{center}
\includegraphics[scale=0.38]{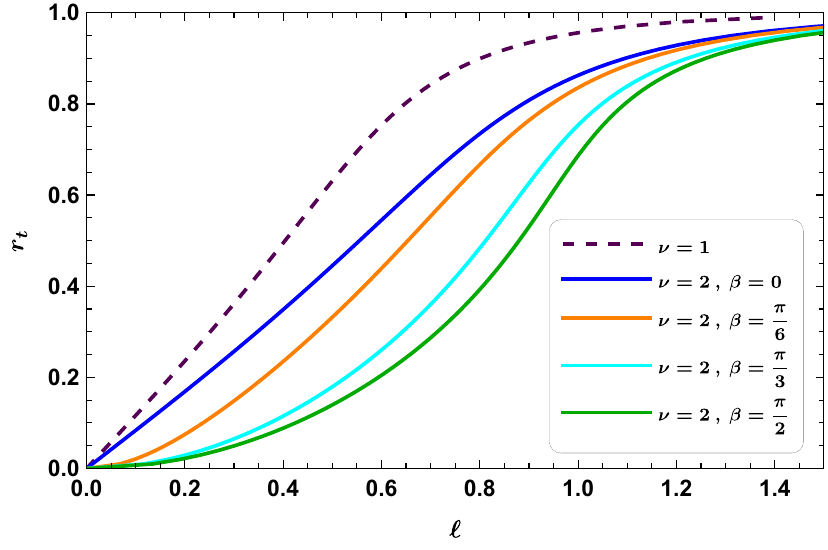}
\hspace*{0.01cm}
\includegraphics[scale=0.38]{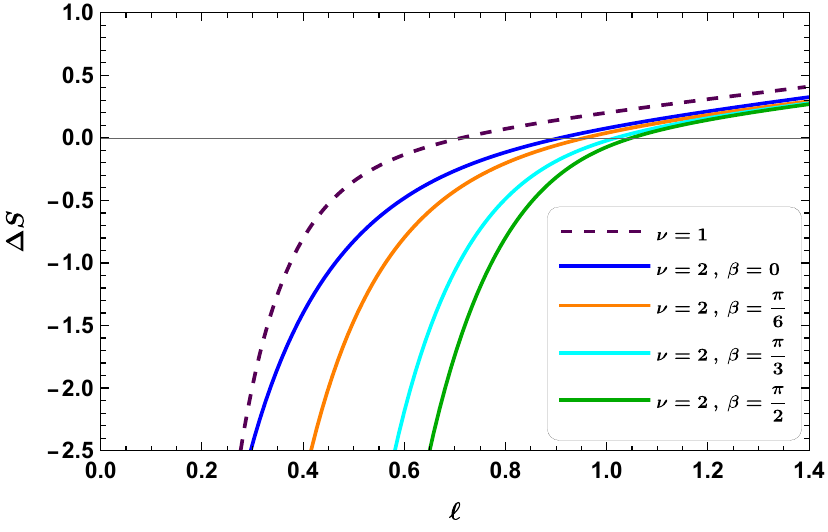}
\hspace*{0.01cm}
\includegraphics[scale=0.38]{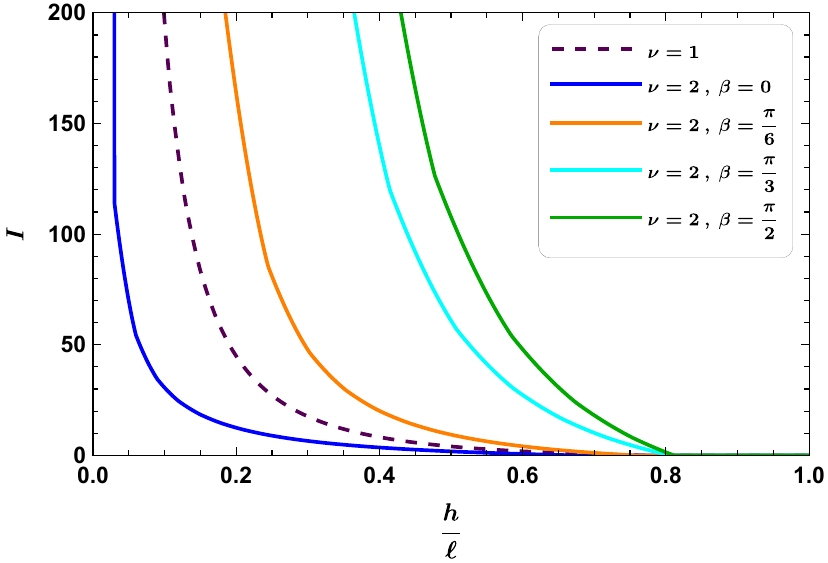}
\end{center}
\caption{\textit{Left}: The turning point of the RT hypersurface as a function of the width of the boundary subregion for different values of the rotation angle. \textit{Middle}: The HEE as a function of $\ell$ for the same values of $\beta$. \textit{Right}: The HMI as a function of $\frac{h}{\ell}$. The solid curves show the anisotropic case with $\nu=2$ and the dashed curve corresponds to isotropic case with $\nu=1$.}
\label{fig:numfig1}
\end{figure}

We present the dependence of the turning point and the corresponding HEE and HMI for specific values of $\beta$ as a function of the width of the subregions and separation between them with $\nu=2$ in figure \ref{fig:numfig1}. Again, the dashed curve corresponds to isotropic case with $\nu=1$. The left panel shows that for a fixed boundary width, as $\nu$ increases from $1$, $r_t$ decreases which means that the bulk potential due to the anisotropy pushes the minimal hypersurface towards the boundary. This behavior is enhanced by increasing the rotation angle from 0 to $\frac{\pi}{2}$. 
%This behavior is enhanced by increasing the rotation angle from 0 to $\frac{\pi}{2}$. The middle panel demonstrates the dependence of the finite part of the HEE defined as $\Delta S\equiv S-S_{\rm div}$ on $\ell$. Note that the divergent part can be read from the asymptotic behavior of the HEE functional near the boundary as follows
%\begin{equation*}
%S_{\rm div}=K^{3/2} \int_{\epsilon} r^{\frac{-2\nu+\sqrt{6}\sqrt{\nu-1}-1}{\nu}}\sqrt{\sin^2\beta+\cos ^2\beta\; G(r)}\, dr,
%\end{equation*}
%where $K$ is a complicated expression which depends on $r_h$ and $\nu$. {\color{red} Also, we must point out that in $\beta=0$ for $\nu \ge 1.67 $, we have an integrable singularity in z=0 \cite{Aref2018}.}
The right panel shows the HMI as a function of the dimensionless boundary quantity $h/\ell$. Based on these plots for fixed $\nu$ we observe that although the HEE decreases with the rotation angle, the HMI increases with $\beta$. This result hold for any value in the range $0\leq \beta \leq \pi/2$.

\begin{figure}[h]
\begin{center}
\includegraphics[scale=0.6]{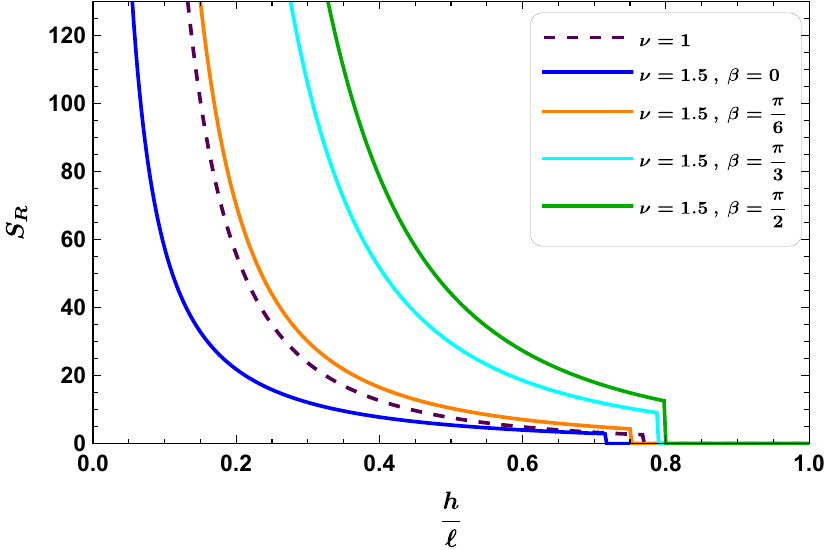}
\hspace*{1cm}
\includegraphics[scale=0.6]{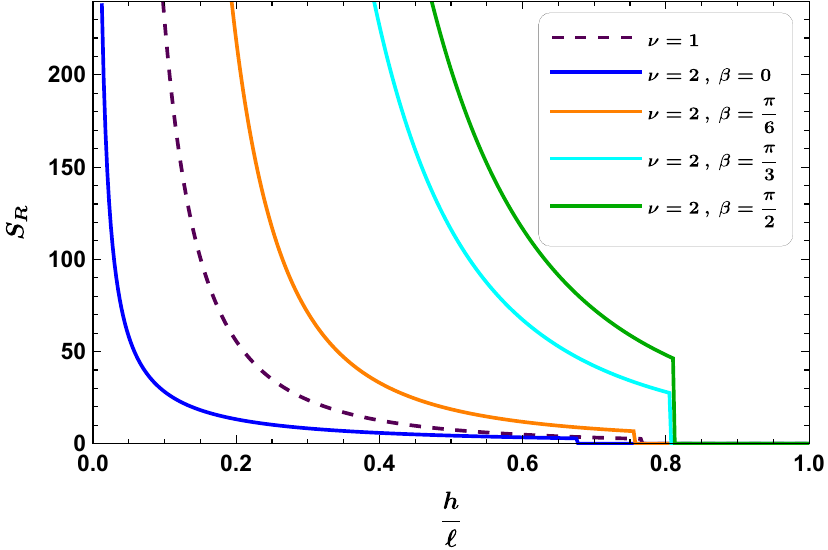}
\end{center}
\caption{Reflected entropy as a function of $\frac{h}{\ell}$ for different values of $\beta$ with $\nu=1.5$ (left) and $\nu=2$ (right). In both plots the dashed curve corresponds to isotropic case with $\nu=1$.}
\label{fig:numfig2}
\end{figure}

\begin{figure}[h]
\begin{center}
\includegraphics[scale=0.7]{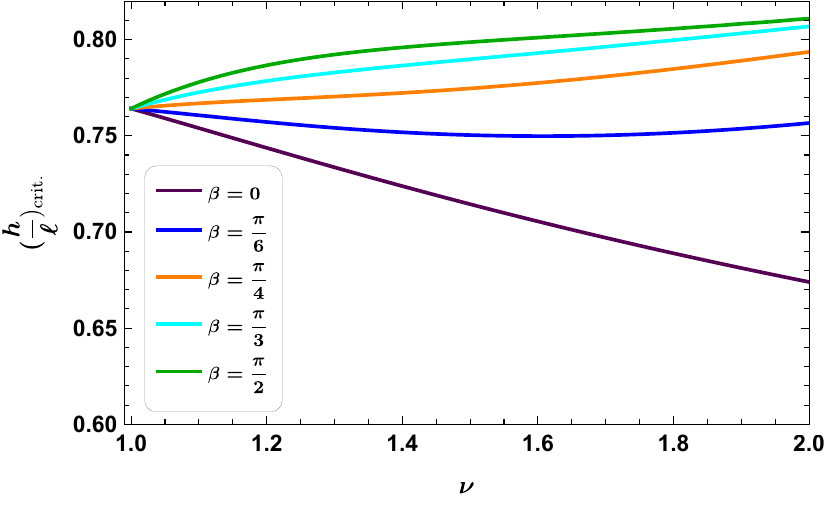}
\end{center}
\caption{Critical separation between the subregions as a function of $\nu$ for different values of $\beta$. For large values of the rotation angle the critical separation is an increasing function of $\nu$ and hence the correlation between the subregions becomes stronger.}
\label{fig:numfig3}
\end{figure}

Figure \ref{fig:numfig2} shows the reflected entropy as a function of $h/\ell$ for different values of $\nu$ and $\beta$.  Let us make a number of observations regarding these numerical results. First, we note that in both plots, in the aforementioned range of the rotation angle, the reflected entropy increases with $\beta$. Next, the phase transition of this measure happens at larger separations between the two subregions comparing to $\beta=0$ case. Hence regarding the reflected entropy as a measure of total correlation between the two subregions, we see that decreasing the rotation angle promotes disentangling between them. Moreover, despite the $\beta=0$ case where the critical separation decreases with $\nu$, for other values of the rotation angles, $\left(\frac{h}{\ell}\right)_{\rm crit.}$ increases with this parameter.  
In figure \ref{fig:numfig3} we present the critical separation as a function of $\nu$ to allow for a meaningful comparison between the different cases. We see that $\left(\frac{h}{\ell}\right)_{\rm crit.}$ becomes a monotonically increasing function of $\nu$ for large values of the rotation angle. For example, if we choose
$\beta=\pi/2$, then  increasing the anisotropy, the critical separation increases which means that the correlation between the subregions becomes stronger. Also for intermediate values of $\beta$, \textit{e.g.}, $\beta=\pi/6$, the critical separation has a minimum at $\nu\sim 1.5$. 

\subsection{Reflected Entropy for large black hole}
\begin{figure}[h]
	\begin{center}
		\includegraphics[scale=0.643]{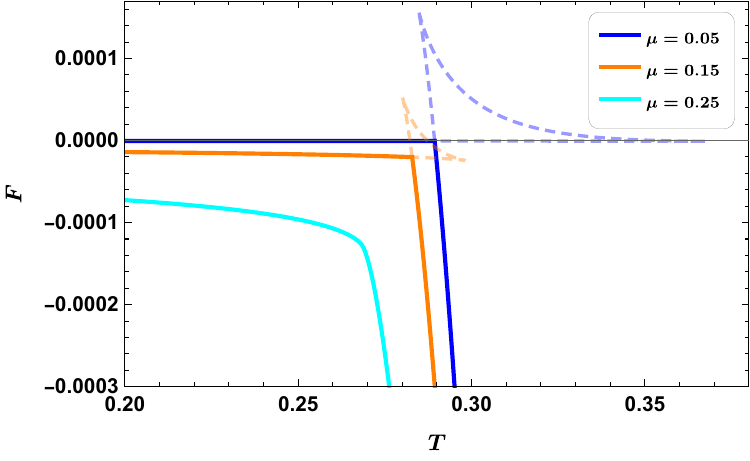}
		\hspace*{0.03cm}
		\includegraphics[scale=0.63]{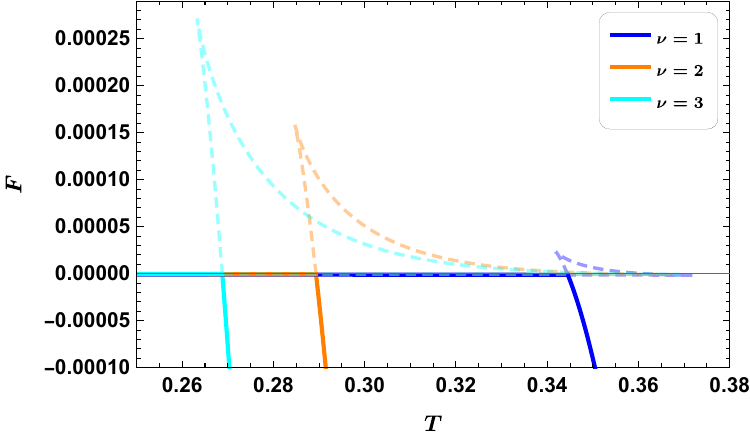}
			\end{center}
	\caption{The free energy as a function of temperature for different values of the chemical potential with $\nu=2$ (left) and different values of the anisotropy parameter with $\mu=0.05$ (right). The dashed region indicates the instability zone and the point where any curve intersects itself corresponds to the small/large black holes phase transition.}
	\label{fig:freeenergy}
\end{figure}

As we have mentioned before, in this geometry the free energy is multivalued and the background exhibits a Van der Waals-like phase transition between small and large black holes.As explained in \cite{Aref2018} for $\mu<\mu_{\rm crit.}$ the curves for $F(T)$ form a swallow tail shape such that an increase in $\mu$ give a decrease in size for the swallow tail region, \textit{e.g.}, see the left panel in figure \ref{fig:freeenergy}. In the right panel we show the same function for different values of the anisotropy parameter with $\mu=0.05$. Interestingly, we see that in this case an increase in $\nu$ give an increase in size for the swallow tail region.  In the large black hole, HEE is multivalued and this phase transition can be realized as a confinement/deconfinement phase transition that depends on the width of the boundary subregion. We also set $r_h=3.7$ throughout this section and hence we are in large black hole phase.
%Some examples of these transitions are shown in figure \ref{fig:freeenergy}. 
 
\begin{figure}[h]
	\begin{center}
		\includegraphics[scale=0.4]{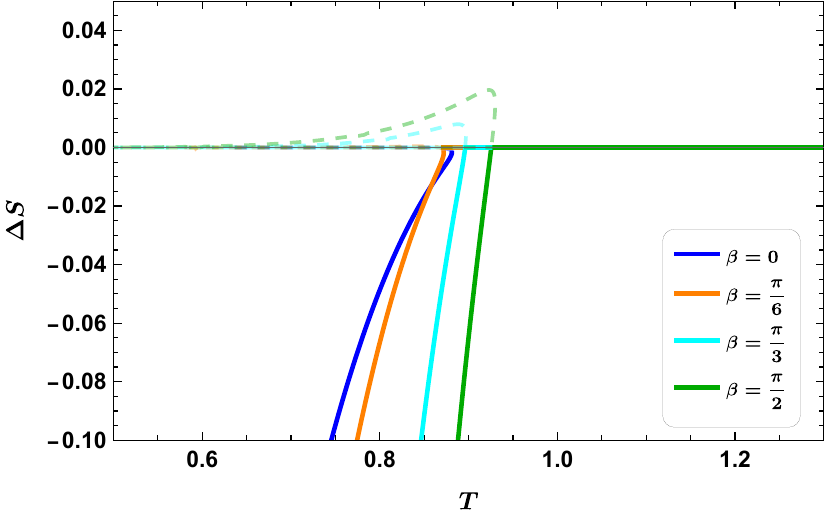}
		\hspace*{0.01cm}
		\includegraphics[scale=0.4]{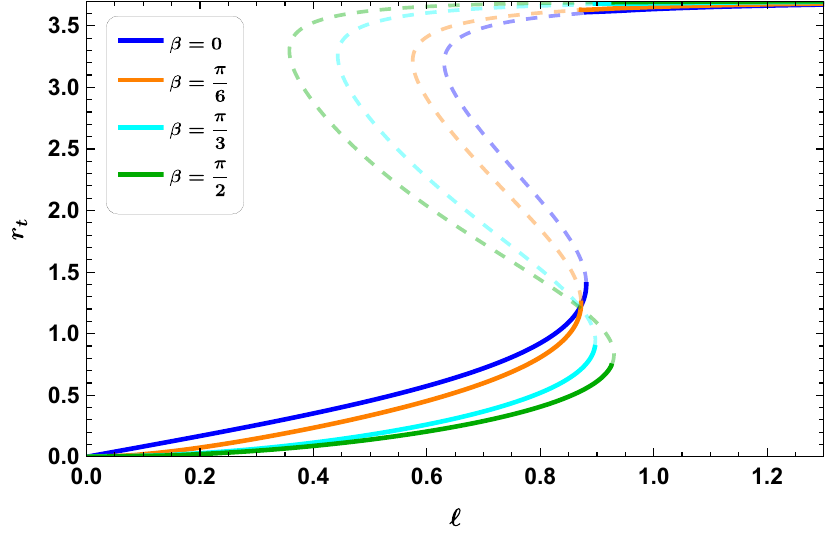}
	\end{center}
	\caption{\textit{Left}: The HEE as a function of $\ell$ for different values of $\beta$. \textit{Right}: The turning point of the RT hypersurface as a function of $\ell$. In all cases the solid curves show the anisotropic case with $\nu=2$ and the dashed curve corresponds to isotropic case with $\nu=1$.}
	\label{fig:numfig4}
\end{figure}

\begin{figure}[h]
	\begin{center}
		\includegraphics[scale=0.4]{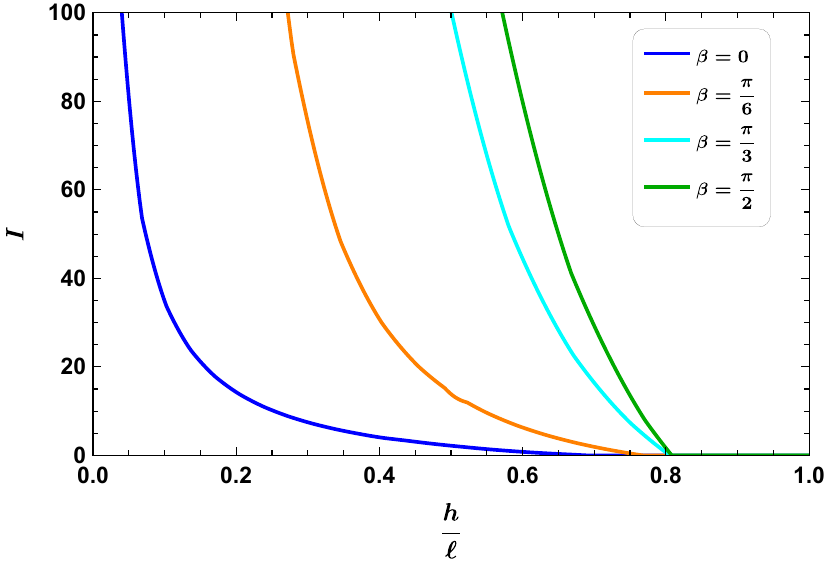}
		\hspace*{0.01cm}
		\includegraphics[scale=0.4]{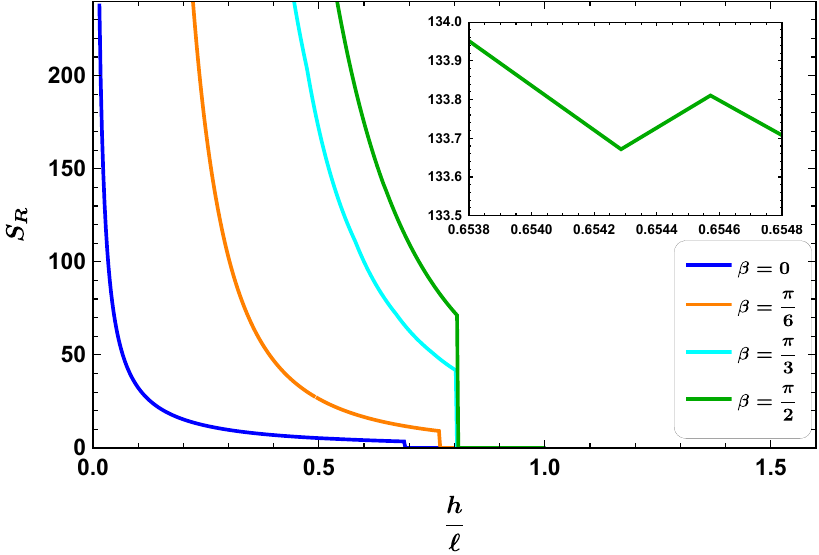}
	\end{center}
	\caption{\textit{Left}: The HEE as a function of $\ell$ for different values of $\beta$. \textit{Middle}: The HMI as a function of $\frac{h}{\ell}$. \textit{Right}: Reflected entropy as a function of $\frac{h}{\ell}$. In all cases the solid curves show the anisotropic case with $\nu=2$ and the dashed curve corresponds to isotropic case with $\nu=1$.}
	\label{fig:numfig4+1}
\end{figure}

Figure \ref{fig:numfig4} shows The turning point of the RT hypersurface and the HEE as a function of width of subregion for $\nu=2$ and different values of $\beta$.
As can be seen, the HEE and the RT hypersurface have multivalued for some width of the subregion and in these widths of the subregion, we must choose the one that has the smallest value of HEE.
Figure \ref{fig:numfig4+1} shows The MI and the Reflected Entropy as a function of $\frac{h}{\ell}$ for $\nu=2$ and different values of $\beta$.
As can be seen in the inset, Due to HEE and the RT hypersurface discontinuity, there is a phase transition in the MI and the Reflected Entropy.

In order to investigate the behavior of these measures during the phase transition, we present the dependence of the turning point and the corresponding HEE and HMI for specific values of $\beta$ as a function of the width of the subregions and separation between them with $\nu=2$ in figure \ref{fig:numfig1}. The left panel shows that for a fixed boundary width, as $\nu$ increases from $1$, $r_t$ decreases which means that the bulk potential due to the anisotropy pushes the minimal hypersurface towards the boundary. This behavior is enhanced by increasing the rotation angle from 0 to $\frac{\pi}{2}$. 
\begin{figure}[h]
	\begin{center}
		\includegraphics[scale=0.5]{m1crsTnu2l1.pdf}
		\hspace*{0.01cm}
		\includegraphics[scale=0.62]{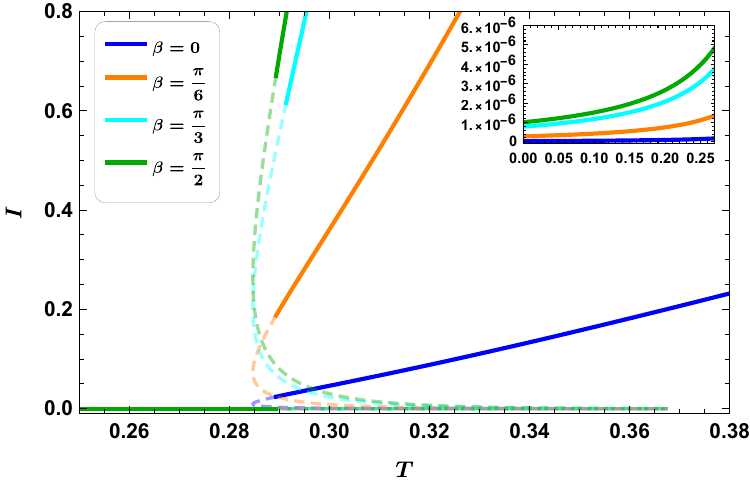}
	\end{center}
	\caption{!!!!!! HEE (left) and HMI (right) as functions of $T$ for different values of $\beta$ with $\ell=1, h=0.2, \nu=2$ and $\mu=0.05$. In both plots the dashed region indicates the instability zone and the inner panels show the same graph with focus on $T\rightarrow 0$ limit. }
	\label{fig:numfig5}
\end{figure}

\begin{figure}[h]
	\begin{center}

		\includegraphics[scale=0.63]{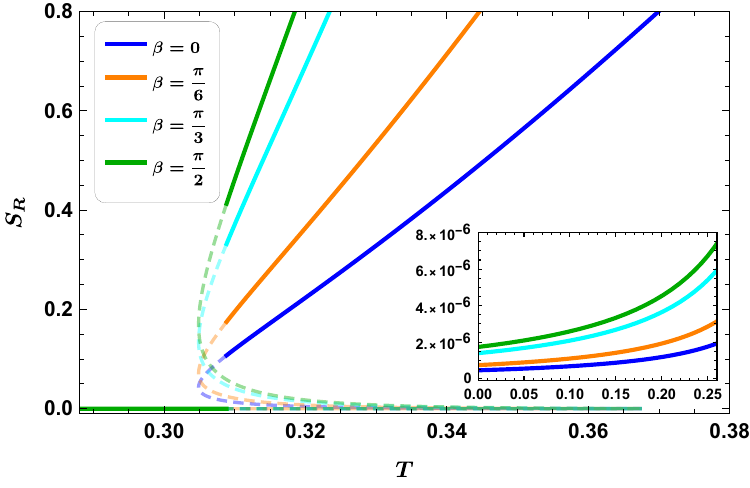}
		\includegraphics[scale=0.63]{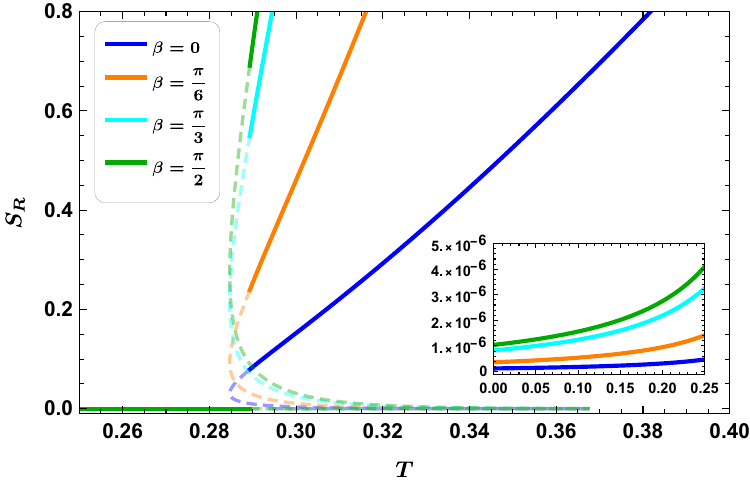}
	\end{center}
	\caption{!!!!!!!!!!!!Reflected entropy as a function of $T$ for different values of $\beta$. Here we set $\ell=1, h=0.2, \mu=0.05$ with $\nu=1.5$ (left) and $\nu=2$ (right). In both plots the dashed region indicates the instability zone and the inner panels show the same graph with focus on $T\rightarrow 0$ limit. }
	\label{fig:numfig6}
\end{figure}

\section{Anisotropic Einstein-Axion-Dilaton Gravities}\label{sec:EAD}
In this section we evaluate reflected entropy and some other holographic entanglement measures for an anisotropic geometry in a family of axion-dilaton gravity theories with the following action \cite{Gubser2008, Nellore2008, Giataganas2018}
\begin{equation}\label{A2}
I=\frac{1}{16\pi G_N}\int \mathrm{d}^5x\sqrt{-g}\left(R-\frac{1}{2}(\partial \phi)^2+V(\phi)-\frac{1}{2}Z(\phi)(\partial\chi)^2\right).
\end{equation}
Here $V(\phi)$ is the dilaton potential and $Z(\phi)$ controls the strength of the coupling between the dilaton and the axion field. As noted in \cite{Giataganas2018}, assuming a linear axion ansatz, \textit{i.e.}, $\chi = a z$,  the equations of motion automatically satisfied such that the underlying geometry breaks isotropy while preserving translation invariance
\begin{equation}\label{metricEAD}
ds^2 =e^{2A(r)}\left(-f(r) dt^2+ dx^2 +dy^2 + e^{2g(r)} dz^2 +\frac{dr^2}{f(r)} \right),\hspace*{1.4cm}\phi=\phi(r).
\end{equation}
The above metric is asymptotically AdS near $r=0$ and $g(r)$ controls the degree of anisotropy between spatial directions. Let us add that, for $V=12$ and $Z=e^{2\phi}$ the dual field theory is conformal. Moreover, a confining boundary theory can be obtained by considering specific dependence for these functions. For instance, choosing
\begin{equation}\label{ADV}
V(\phi)=12\cosh(\sigma\phi)+b\phi^2,\qquad Z(\phi)=e^{2\gamma\phi},
\end{equation}
with $ b \equiv \frac{\Delta(4 - \Delta)}{2} - 6 \sigma^2$, the corresponding boundary theory has a confined phase for $ \sigma \ge \sqrt{2/3} $\cite{Gursoy:2007er}. Here $\Delta$ is the scaling dimension of the scalar operator dual to $\phi$. In the following, we study influence of anisotropy on holographic information measures in different backgrounds.

\subsection{Non-conformal Boundary Theory}\label{sec:nonconformal}
In this case we consider a marginal scalar operator with $\Delta=4$ at zero temperature, \textit{i.e}, $f(r)=1$. A perturbative solution for the equations of motion in the small aniotropy limit was found in \cite{Ghasemi2019} where the metric is given by \eqref{metricEAD} with 
\begin{align}\label{pertmetric}
A(r)=&-\log(r)-\frac{a^2r^2}{72}+\frac{a^4r^4}{1200}(3\gamma^2+1)(1-5\log(ar))+\mathcal{O}(ar)^6,\\
g(r)=&\frac{a^2r^2}{8}-\frac{a^4r^4}{2592}\left(31+81\gamma^2-54(3\gamma^2+1)\log(ar)\right)+\mathcal{O}(ar)^6.
\end{align}
Let us mention that in this background, the $xy$ plane is isotropic and hence the rotation of the strip around the $z$ axis has no effect on holographic correlation measures. Further, in comparing \eqref{metricEAD}  with metric \eqref{genmetric}, we should identify
\begin{align}
H(r)=r^2 e^{2 A(r)},\qquad G_1(r)=G_2(r)=1, \qquad G_3(r) = e^{2g(r)},
\end{align}
and thus
\begin{align}
\mathcal{G}=e^{2g(r)} , \qquad \mathcal{T}(r,\beta)=\sin^2 \beta + e^{2g(r)} \cos^2 \beta.
\end{align}
The corresponding expression for HEE and reflected entropy can be obtained using eqs. \eqref{GS} and \eqref{GEW} and the above identifications. In the following, we first provide a numerical analysis and examine the dependence of different measures on the the anisotropy parameter. Next, we carry out a perturbative
analysis for calculating these measures in the specific regimes of the parameter space. 

\subsubsection*{4.1.1 Numerical Results}

\begin{figure}[h]
	\centering
	\includegraphics[width=0.45\linewidth]{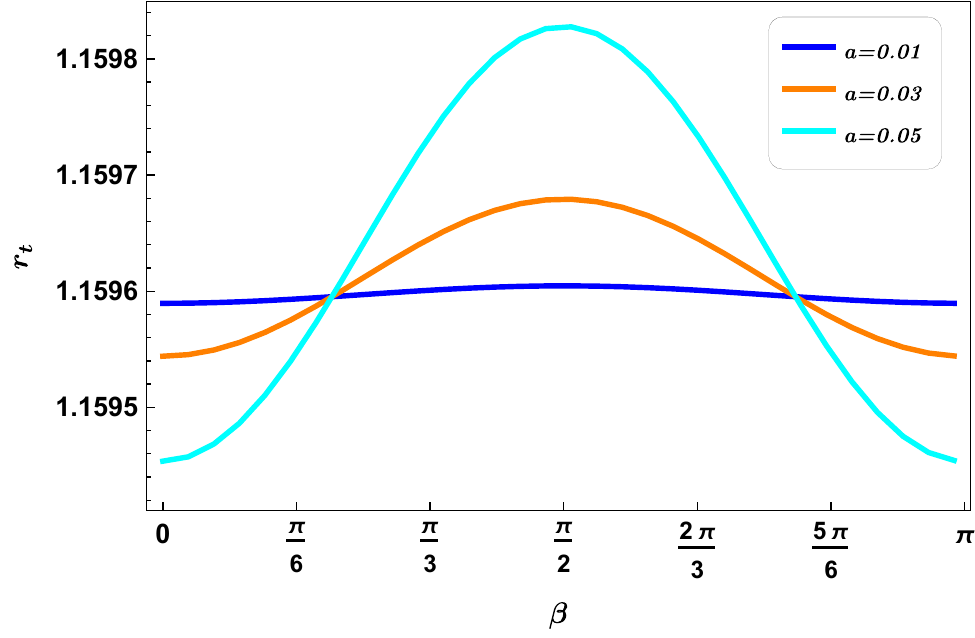}
	\hspace*{1cm}
	\includegraphics[width=0.47\linewidth]{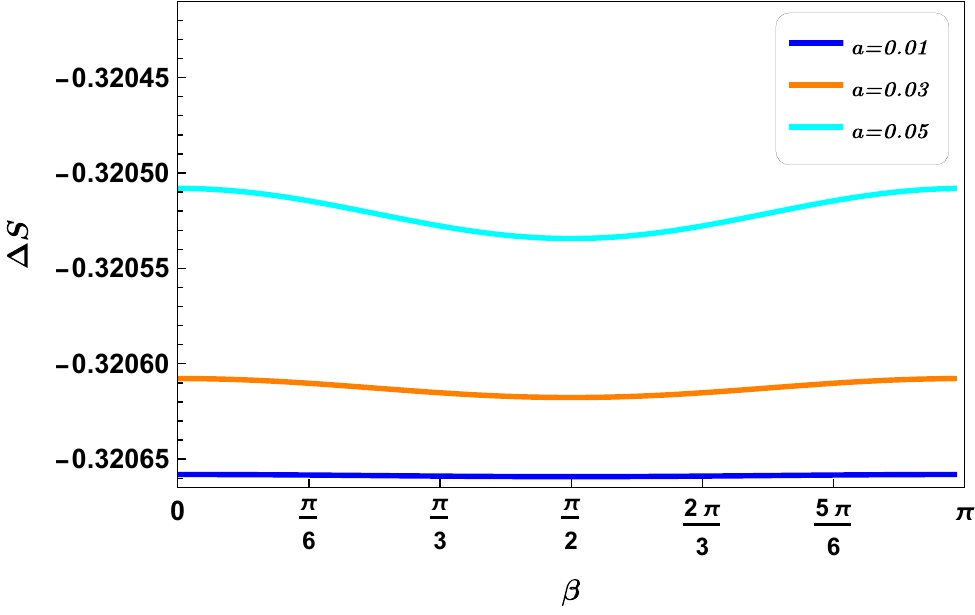}
		\caption{The turning point of the RT hypersurface (left) and HEE (right) as functions of the rotation angle for different values of the anisotropy parameter. Here we set $\ell=1$.}
	\label{fig:EADrt}
\end{figure}

In fig.\ref{fig:EADrt} we show the turning point of the RT hypersurface and HEE as functions of the rotation angle for different values of the anisotropy parameter for a fixed extent of the boundary subregion. The left panel shows that in a specific range of the rotation angle, \textit{i.e.}, $\frac{\pi}{6}\lesssim\beta\lesssim\frac{5\pi}{6}$, increasing the anisotropy, the RT hypersurfaces reach deeper into the bulk, so they carry more information about the geometry. Notice that validity of the background solution was assumed and we set the subleading terms in eq. \eqref{pertmetric} to zero. This asymptotic behavior is valid for $ar_t\ll 1$, or equivalently, $a\ell\ll 1$, the range of anisotropy that we shall consider in the following. The right panel illustrates the HEE, which is regularized by subtracting the divergent part of eq. \eqref{GS}. This divergent term up to the order $a^2$ correction becomes
\begin{equation}\label{sdiv}
S_{\rm div}=\frac{1}{2G_N} \left(\frac{1}{ \epsilon ^2}-\frac{1}{24} a^2 (1+3 \cos (2 \beta )) \log \epsilon\right).
\end{equation}

\begin{figure}[h]
	\centering
	\includegraphics[width=0.47\linewidth]{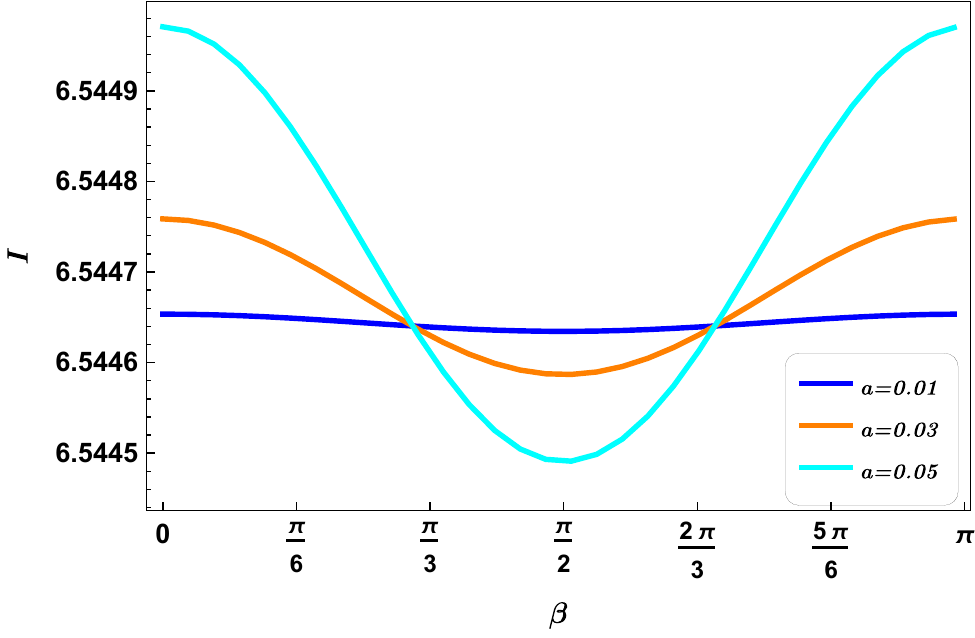} \qquad
	\includegraphics[width=0.47\linewidth]{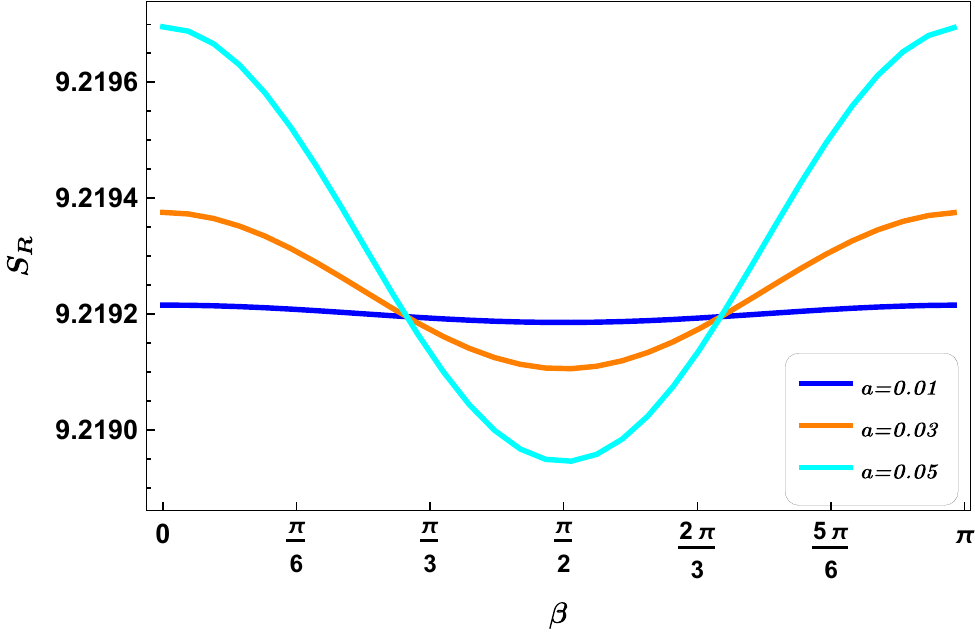}
	\caption{ HMI (left) and reflected entropy (right) as functions of $\beta$ for $\ell=1, h=0.2$ and different values of $a$.}
	\label{fig:EADS}
\end{figure}

\begin{figure}[h]
	\centering
	\includegraphics[width=0.5\linewidth]{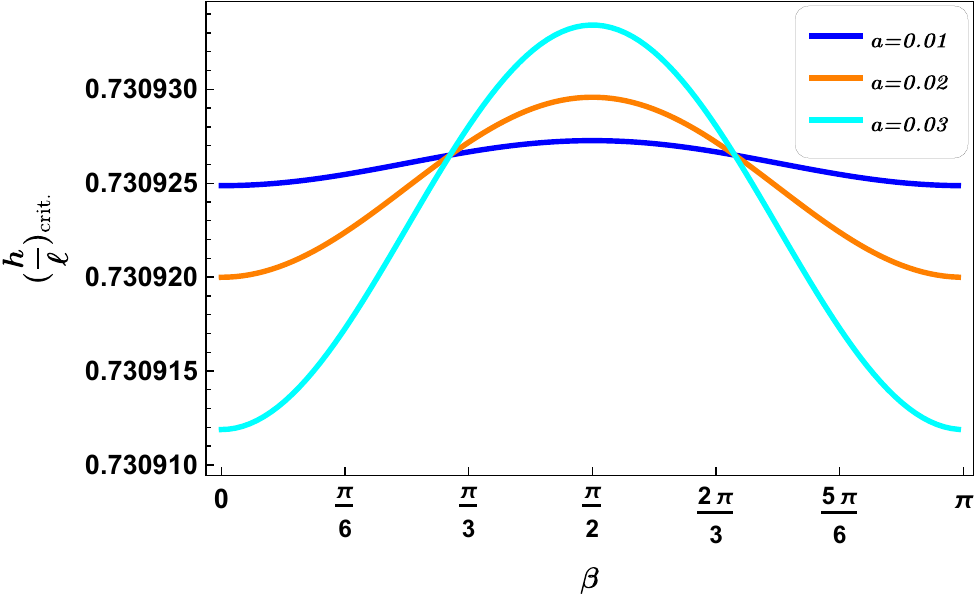}
	\caption{Critical separation between the subregions as a function of $\beta$ for different values of $a$. For $\beta\sim \pi/2$ of the rotation angle the critical separation is an increasing function of the anisotropy parameter and hence the correlation between the subregions becomes stronger.}
	\label{fig:EADPT}
\end{figure}

In fig.\ref{fig:EADS} we show the HMI and reflected entropy as functions of $\beta$ for different values of $a$ and specific values of $\ell$ and $h$. Clearly the qualitative behavior of these two measures are similar as expected. As we mentioned before both the HMI and reflected entropy are measures of total correlation between subregions, hence the holographic calculations reproduce the expected behavior. Moreover, based on these figures, the corresponding correlations develop a minimum at $\beta=\pi/2$.
In  fig.\ref{fig:EADPT}, we show the phase transition point of reflected entropy as a function of the rotation angle. Interestingly, we see that for $\frac{\pi}{6}\lesssim\beta\lesssim\frac{5\pi}{6}$, increasing the anisotropy, the critical separation increases which means that the correlation between the subregions becomes stronger. We will confirm these observations as well as some new results with a perturbative analysis below.

\subsubsection*{4.1.2 Perturbative Treatment}
As we mentioned before in $a\ell\ll 1$ limit the metric \eqref{metricEAD} is a small deformation of pure
AdS, thus we can use a perturbative expansion to compute the variation of holographic information measures. To do so, we can perform a change of variables in the corresponding expressions for $\ell, S$ and $S_R$ to the dimensionless coordinate $u=\frac{r}{r_t}$. In this situation, the corresponding boundary quantities become  
\begin{eqnarray}
\ell &=&2 r_t\int_{0}^{1} \frac{\sqrt{\mathcal{T}(u,\beta) }}{e^{g(u)}\sqrt{\frac{e^{2g(u)}e^{6A(u)}}{e^{2g(1)}e^{6A(1)}}-1}} du,\label{L2}\\
S&=&\frac{L^2r_t}{2 G_N} \int_{\epsilon/r_t}^{1} \frac{e^{g(u)}e^{6A(u)}\sqrt{\mathcal{T}(u,\beta) }}{\sqrt{e^{2g(u)}e^{6A(u)}-e^{2g(1)}e^{6A(1)}}}du,\label{EE2}\\
S_R&=&\frac{L^2r_t}{2 G_N } \int_{r_d/r_t}^{r_u/r_t} e^{3A(u)}\sqrt{  \mathcal{T}(u,\beta)  } du.\label{RE2}
\end{eqnarray}
Now we expand eq. \eqref{L2} in the limit $a\ell\ll 1$ to find the leading corrections to $r_t$ compared to its
pure AdS value. Note that in this case the corresponding turning point is close to the boundary, \textit{i.e.}, $ar_t\ll 1$. In this limit  eq. \eqref{L2} can be written in terms of the following expansion
\begin{equation}
\ell=2	r_t \int_0^1 \frac{u^3}{\sqrt{1-u^6}} \, du  +2	r_t^3 \ a^2 \int_0^1      \frac{  \left(2 \cos ^2(\beta )-\left(3 \left(u^6+u^4+u^2\right)-2\right) \sin ^2(\beta )\right)}{24 \sqrt{\frac{1}{u^6}-1} \left(u^4+u^2+1\right)}  \, du.  
\end{equation}
The above integral can be evaluated explicitly yielding
\begin{equation}\label{m2l}
\ell=2 r_t \left(c+a^2 r_t^2 \mathcal{C}_1 (\beta) \right),
\end{equation}
where
\begin{equation}
c=\frac{\sqrt{\pi } \Gamma \left(\frac{2}{3}\right)}{\Gamma \left(\frac{1}{6}\right)},\qquad\qquad \mathcal{C}_1(\beta)=-0.005+0.021 \cos2\beta.
\end{equation}
Notice that the first term in eq. \eqref{m2l} is the pure AdS contribution. Inverting this equation, we can represent the turning point as
a function of $\ell$
\begin{equation}\label{m2rt}
r_t	=\frac{\ell}{2c} \left(1-a^2 \ell^2 \frac{\mathcal{C}_1(\beta)}{4\,c^3}\right).
\end{equation}
Let us comment on the properties of the above result: First, we observe that the location of the turning point is unaffected for $\beta_1\sim 0.66$ (or equivalently $\beta_2\sim \pi-0.66$) where $\mathcal{C}_1(\beta_{1, 2})=0$. Moreover, for $\beta_1\leq \beta \leq\beta_2$, $\mathcal{C}_1(\beta)$ is negative and therefore the correction to $r_t$ in eq. \eqref{m2rt} is positive. Hence the RT hypersurfaces can probe more of the bulk geometry due to the presence of anisotropy. These results are consistent with the previously numerical results illustrated in the left panel of figure \ref{fig:EADrt}.

Now we proceed to examine the leading correction to HEE from eq. \eqref{EE2} using a similar reasoning to that above. Let us mention that it will be more convenient to separate the divergent piece in this Integral. A simple analysis shows that in this case the HHE takes the following form 
%To do so, we rewrite the HEE functional as follows 
%\begin{align}\label{sx}
%	S=&\frac{L^{2}}{G_N}r_t\int_{0}^{1}\mathrm{d}u\left(\frac{1}{u^3 \sqrt{u^6-1}}-\frac{1}{u^3}+\frac{a^2 \cos ^2(\beta )}{r_t} \frac{\left(2 u^4+u^2+1\right) \sqrt{-\frac{1}{u^6-1}}}{12 \left(u^5+u^3+u\right)}-\frac{2}{24 u}\right. \notag\\
%	&\qquad\qquad\qquad\qquad\left.+\frac{a^2 \sin ^2(\beta )}{r_t}\frac{\left(u^4-u^2-1\right) \sqrt{-\frac{1}{u^6-1}}}{24 \left(u^5+u^3+u\right)}-\frac{1}{24 u}\right)\notag\\
%	+&\frac{L^{2}}{G_N}r_t\int_{\epsilon/r_t}^{1}\mathrm{d}u\left(\frac{1}{r_t^3 u^3}+\frac{a^2 \left(2 \cos ^2(\beta )-\sin ^2(\beta )\right)}{24 r_t u}\right).
%\end{align}
\begin{equation}\label{HEEexpand}
S=\frac{1}{4G_N} \left(\frac{1}{\epsilon ^2}-\frac{c}{r_t^2}\right)+\frac{1}{2G_N}a^2\left(\mathcal{C}_2(\beta)\log\frac{r_t}{\epsilon }+ \mathcal{C}_3(\beta)\right)+\mathcal{O}(a^4),
\end{equation}
where
\begin{equation}
\mathcal{C}_2(\beta)=\frac{1}{48}\left(1+3\cos 2\beta\right),\qquad      \mathcal{C}_3(\beta)=0.021+0.014 \cos2\beta.
\end{equation}
In principle then, we can invert the above expressions to write our result in terms of the width of the 
entangling region. Combining Eqs. \eqref{m2rt} and \eqref{HEEexpand}, we obtain the first order
correction to HEE as follows
\begin{equation}\label{anaHEE}
S=\frac{1}{4G_N}\left(\frac{1}{ \epsilon ^2}-\frac{4c^3}{\ell^2}\right)+\frac{1}{2G_N}a^2\left(\mathcal{C}_2(\beta)  \log \frac{\ell}{2 c\epsilon}-\mathcal{C}_1(\beta)+\mathcal{C}_3(\beta)\right).
\end{equation}
A key feature of the above result is the appearance of a new universal logarithmic term which depends on the anisotropy parameter. The coefficient of this term depends also on the rotation angle of the entangling region such that in the $\beta\sim 0.955$ limit where $\mathcal{C}_2(\beta)=0$, vanishes. Roughly, we can think of this universal term as characterizing when the isotropy is broken in the underlying boundary theory. Similarly, as shown in \cite{RezaMohammadiMozaffar:2016lbo}, if instead we choose a background which breaks the translation invariance the structure of the universal terms of HEE is modified. Next, the HMI can be determined using eqs. \eqref{HMI} and \eqref{anaHEE} as follows
\begin{equation}
I=\frac{c^3}{G_N}\left(-\frac{2}{\ell^2}+\frac{1}{h^2}+\frac{1}{(2\ell+h)^2}\right)+\frac{1}{2G_N}a^2\mathcal{C}_2(\beta)\log\frac{\ell^2}{h(2\ell+h)}.
\end{equation}

Finally expanding eq. \eqref{RE2} we can derive the following expression for the reflected entropy at leading order
\begin{equation}
S_R =\frac{1}{4 G_N} \left(\frac{1}{r_d^2}-\frac{1}{r_u^2}\right)+\frac{1}{2 G_N}a^2 \mathcal{C}_2(\beta) \log \frac{r_u}{r_d}.
\end{equation}
We can use eq. \eqref{m2rt} to rewrite the above result as follows
\begin{equation}\label{srpertnonconformal}
S_R =\frac{c^2}{G_N} \left(\frac{1}{h^2}-\frac{1}{(2\ell+h)^2}\right)+\frac{1}{2 G_N}a^2 \mathcal{C}_2(\beta) \log \frac{2\ell+h}{h}.
\end{equation}
Interestingly, we see that for $\mathcal{C}_2(\beta)=0$, where the universal term vanishes, the reflected entropy is independent of the anisotropy parameter (see fig. \ref{fig:EADS}). In this case the corresponding transition point of the HMI and reflected entropy is independent of $a$ which is consistent with the results presented in fig. \ref{fig:EADPT}.

\subsection{Strongly Coupled Anisotropic Plasma}\label{secplasma}

In this section we extend our analysis to another  five-dimensional axion-dilaton-gravity theory which is dual to a strongly coupled anisotropic plasma at finite temperature. The corresponding action and dilaton potential are given by eqs. \eqref{A2} and \eqref{ADV} with $\sigma=0$. Again, we consider a linear axion ansatz, \textit{i.e.}, $\chi=a z$. As shown in \cite{Mateos:2011tv}, in high-temperature limit, it is possible to find analytic expressions for the metric as follows
\begin{equation}
ds^2=\frac{e^{-\frac{\phi(r)}{2}}}{r^2}\left(-{f}(r)b(r)dt^2+dx^2+dy^2+e^{-\phi(r)}dz^2+\frac{dr^2}{{f}(r)}\right),
\end{equation}
where 
\begin{align}
f(r)=1-\frac{r^4}{r_h^4}+\frac{a^2}{24r_h^2}\left(8r^2(r_h^2-r^2)-10r^4\log2+(3r_h^4+7r^4)\log\left(1+\frac{r^2}{r_h^2}\right)\right),\\
b(r)=1-\frac{a^2r_h^2}{24}\left(\frac{10r^2}{r_h^2+r^2}+\log\left(1+\frac{r^2}{r_h^2}\right)\right),\hspace*{1cm}\phi(r)=-\frac{a^2r_h^2}{4}\log\left(1+\frac{r^2}{r_h^2}\right).
\end{align}
By high-temperature limit, we mean that $a\ll T$ which implies that $ar_h\ll 1$.

\begin{figure}[h]
	\centering
	\includegraphics[width=0.45\linewidth]{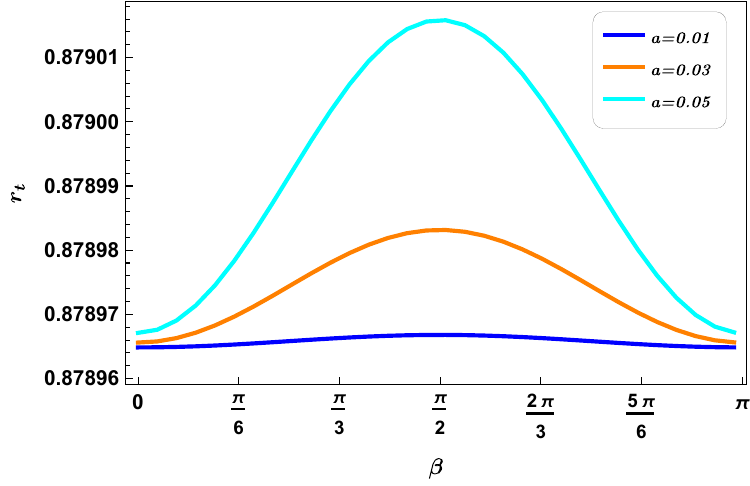}
	\hspace*{1cm}
	\includegraphics[width=0.47\linewidth]{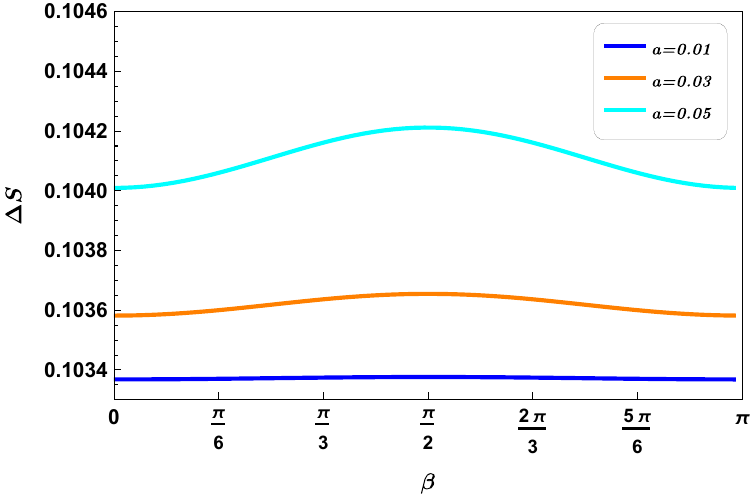}
		\caption{The turning point of the RT hypersurface (left) and HEE (right) as functions of the rotation angle for different values of the anisotropy parameter. Here we set $\ell=1$.}
	\label{fig:plasmarts}
\end{figure}

\begin{figure}[h]
	\centering
	\includegraphics[width=0.47\linewidth]{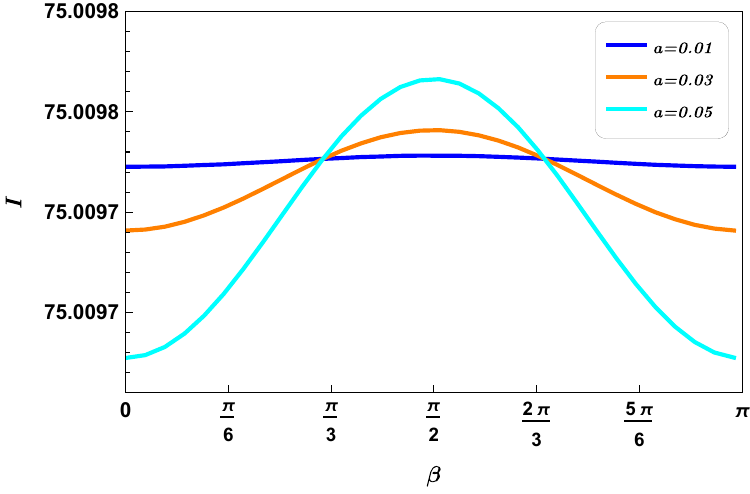}
		\hspace*{0.7cm}
	\includegraphics[width=0.47\linewidth]{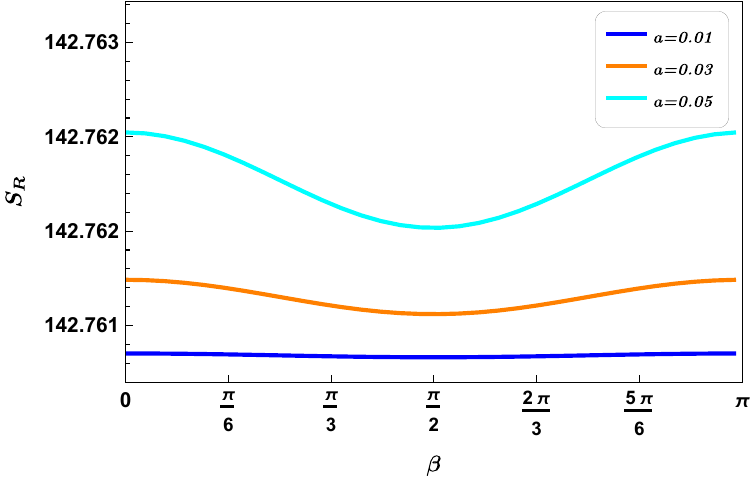}
	\caption{ HMI (left) and reflected entropy (right) as functions of $\beta$ for $\ell=0.1, h=0.05, r_h=1$ and different values of $a$.}
	\label{fig:plasmaIRE}
\end{figure}

The corresponding analysis for evaluating the holographic measures follows similarly to the previous section, with the obvious replacement of the metric components in eqs. \eqref{GS} and \eqref{GEW}. Unfortunately, it is not possible to compute the dependence of the measures on $a$ perturbatively even for certain values of the rotation angle. Thus, in what follows we just present the numerical results. Let us add that a simple analysis shows that in this case the divergent term of the HEE is the same as eq. \eqref{sdiv}. For simplicity, we set $r_h=1$ throughout the following. To illustrate the numerical results, we show the holographic measures as functions of the rotation angle for different values of the anisotropy parameter in figures \ref{fig:plasmarts} and \ref{fig:plasmaIRE}.  

The left panel in \ref{fig:plasmarts} shows the turning point of the RT hypersurface for $\ell=1$. Clearly, increasing the anisotropy, the RT hypersurfaces reach deeper into the bulk, thus they carry more information about the geometry. The right panel illustrates the finite part of the HEE which is an increasing function of $a$. Figure \ref{fig:plasmaIRE} shows the HMI and reflected entropy for specific values of $\ell$ and $h$. Although the reflected entropy increases with the anisotropy parameter for all values of the rotation angle, HMI is not a monotonic function of $a$. Moreover, at $\beta=\pi/2$ the HMI becomes maximum where the reflected entropy developes a minimum. Interestingly, while both HMI and reflected entropy are measures of total correlation between subregions, they do not behave in the same manner in this anisotripic boundary state. This behavior contrast with the results depicted in figure \ref{fig:EADS}, where these measures behave in the same manner in a non-conformal boundary theory. We do not fully understand what is the reason for this behavior and leave it for future study. 

\begin{figure}[h]
	\centering
	\includegraphics[width=0.45\linewidth]{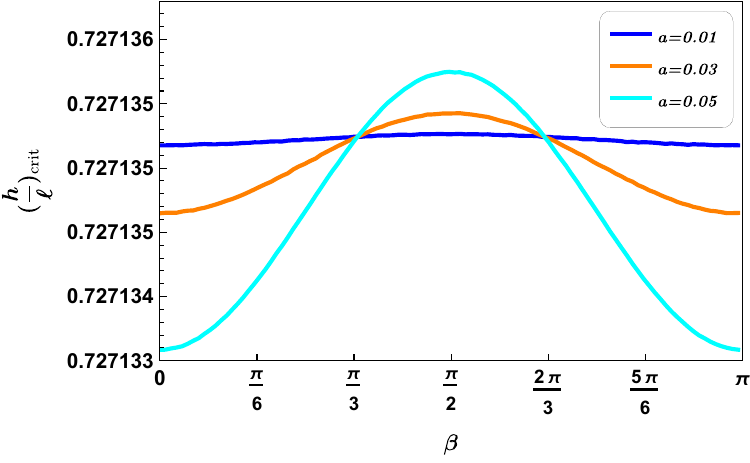}
	\hspace*{1cm}
	\includegraphics[width=0.45\linewidth]{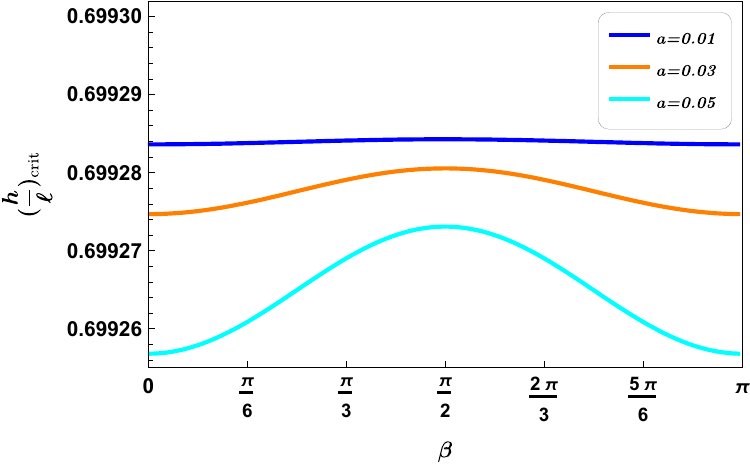}
	\caption{left: HMI Phase transition point as a function of $\beta$ for different a, $r_h=1$ and  $\ell=0.1$.Right: HMI Phase transition point as a function of $\beta$ for different a, $r_h=1$ and  $\ell=0.4$.}
	\label{fig:2EADPT}
\end{figure}

\section{Lifshitz-like Anisotropic Models}\label{sec:Lifshitz}
In this section we extend our studies to a specific geometry with anisotropic Lifshitz scale invariance first studied in \cite{Azeyanagi2009}. This geometry is a type IIB supergravity solution and generated by intersections of D3 and D7 branes. The corresponding metric is given by
\begin{equation}\label{Lifmetric}
ds^2=\frac{\tilde{R}^2}{r^2}\left(-f(r)dt^2+\frac{dr^2}{f(r)}+dx^2+dy^2+r^{\frac{2}{3}}dz^2\right),\hspace*{2cm}f(r)=1-\mu \,r^{\frac{11}{3}},
\end{equation}
where $\tilde{R}^2=\frac{11}{12}R^2$ is the curvature radius of the spacetime. Further, $\mu$ gives the mass parameter of the black brane. This geometry is dual to a nonrelativistic boundary theory with the following expressions for temperature and energy density, respectively
\begin{eqnarray}\label{tempLif}
T=\frac{11}{12\pi}\mu^{\frac{3}{11}},\hspace*{2cm}\varepsilon=\frac{\tilde{R}^{3}}{6\pi G_N}\mu.
\end{eqnarray}
We see that for $\mu=0$ metric \eqref{Lifmetric} is invariant under an anisotropic scaling transformation $(t, r, x, y, z) \rightarrow (\lambda t, \lambda r, \lambda x, \lambda y, \lambda^{2/3} z)$ and thus can be regarded as a gravity dual of Lifshitz-like fixed point with dynamical exponent $\xi=\frac{3}{2}$. Let us recall that different aspects of holographic probes including viscosities and HEE in this model has been studied in \cite{Azeyanagi2009}. Note that in this geometry, the strength of anisotropy between spatial directions is fixed, thus we only study the $\beta-$dependence of reflected entropy. In comparing the above background with metric \eqref{genmetric}, we should identify
\begin{equation}\label{Liffunc}
G_1(r)=G_2(r)=H(r)=1,\qquad\qquad G_3(r)=r^{\frac{2}{3}}.
\end{equation}
The corresponding expression for HEE and reflected entropy can be obtained using eqs. \eqref{GS} and \eqref{GEW} and the above identifications. Before examining the full $\beta-$dependence of correlation measures, we would like to study the structure of divergent terms. Notice that because the metric \eqref{Lifmetric} is not asymptotically AdS, the corresponding divergent terms that appear in HEE are more complicated. A straightforward calculation for $\beta>0$, yields the following \footnote{For $\beta=0$ we have $S_{\rm div}\sim \epsilon^{-5/3}$.}
\begin{equation}\label{sdivLif}
S_{\rm div}=\frac{L^2\sin\beta}{4G_N}\left(\frac{1}{\epsilon^2}+\frac{3\cot^2\beta}{4\epsilon^{4/3}}-      \frac{3\cot^4\beta}{8\epsilon^{2/3}}-\frac{\cot^6\beta}{8}\log\epsilon\right).
\end{equation}
Again, we see that a new universal logarithmic term appears, whose coefficient depends on the rotation
angle. 

Unfortunately, it is not possible to find the behavior of the reflected entropy analytically for general $\beta$. In the following, we present a combination of numerical and analytic results on the behavior of correlation measures for strip shaped boundary subregions. First, we provide a numerical analysis and
examine the various properties of reflected entropy as a function of $\beta$. Next, we will show
that at zero temperature and for specific values of the rotation angle, $\Sigma_{A\cup B}$ is a geodesic whose length can be expressed analytically in closed form, which enables us to directly extract its scaling behavior as a function of $h$ and $\ell$. We also carry out a perturbative analysis to compute low temperature corrections to reflected entropy at leading order.

\subsection{Numerical Results}
\begin{figure}[h]
	\centering
	\includegraphics[width=0.45\linewidth]{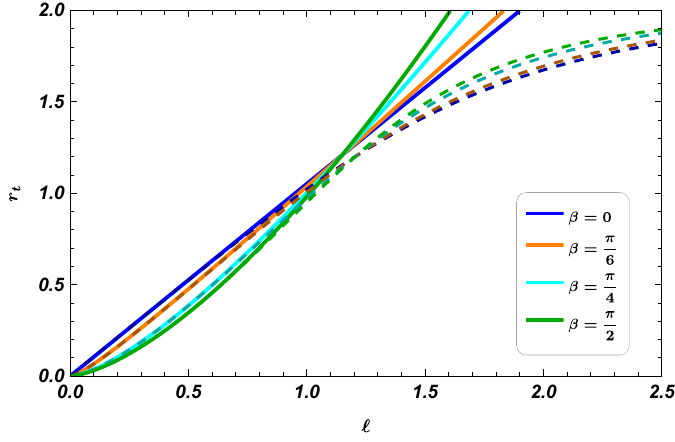}\hspace*{1cm}
	\includegraphics[width=0.45\linewidth]{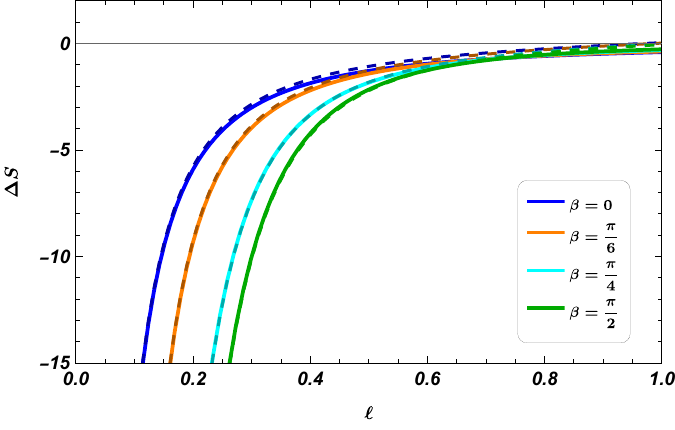}
	\caption{The turning point of the RT hypersurface (left) and HEE (right) as functions of $\ell$ for different values of the rotation angle. The dashed curves represent the finite temperature results and the solid curves correspond to $T=0$ case.}
	\label{fig:LifrtS}
\end{figure}

In fig.\ref{fig:LifrtS} we show the turning point and the finite part of HEE as functions of $\ell$ for several values of the rotation angle. In the figure, the dashed curves represent the finite temperature results and the solid curves correspond to $T=0$ case. According to the left panel, for zero temperature case, at $\ell_c\sim 1.15$ different curves coincide and the location of the turning point is independent of $\beta$. Further, we note that for small subregions, \textit{i.e.,} $\ell<\ell_c$, the turning point decreases in anisotropic case compared to its AdS value which means that the bulk potential due to the anisotropy pushes the RT hypersurface towards the boundary. This behavior is enhanced by increasing the rotation angle from $0$ to $\frac{\pi}{2}$. Moreover, from the right panel we see that for small subregions the finite part of the HEE is a monotonically decreasing function of $\beta$. 
%We can estimate the point where the monotonically decreasing behavior of HEE changes into monotonically increasing as $\ell\lesssim 0.74$. 

\begin{figure}[h]
	\centering
	\includegraphics[width=0.45\linewidth]{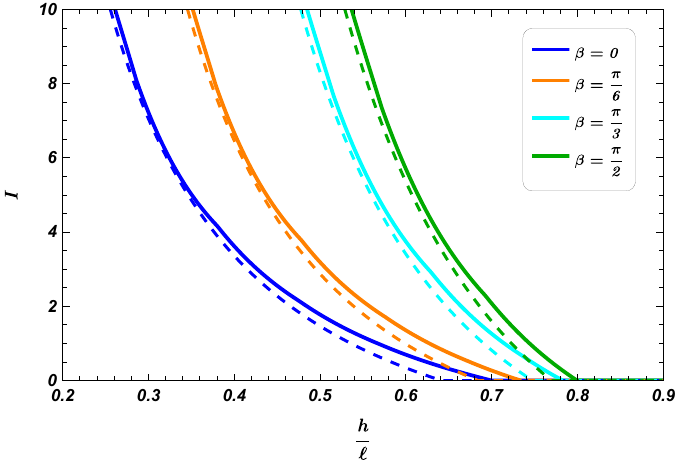}\hspace*{1cm}
	\includegraphics[width=0.45\linewidth]{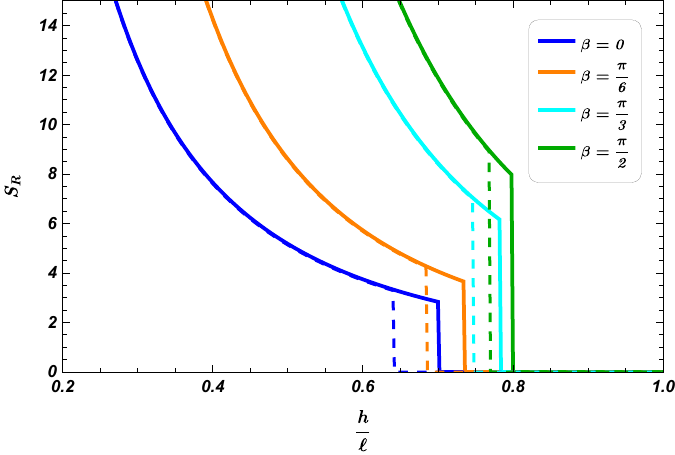}
	\caption{HMI (left) and reflected entropy (right) as functions of $\frac{h}{\ell}$ for different values of $\beta$.}
	\label{fig:LifMIEW}
\end{figure}

Figure \ref{fig:LifMIEW} shows the HMI and reflected entropy as functions of $\frac{h}{\ell}$ for different values of $\beta$. Let us make a number of observations about these numerical results. First, we note that both HMI and reflected entropy are monotonically increasing functions of $\beta$. Next, the phase transition of the reflected entropy happens at larger separations between the subregions comparing to $\beta=0$ case. Hence regarding the reflected entropy as a measure of total correlation between the subregions, we see that decreasing the rotation angle promote disentangling between them. Further, turning on the temperature, the phase transition of reflected entropy happens at smaller separations between the two subregions comparing to $T=0$ case.
Thus the thermal excitations decrease the total correlation between the subregions as expected.

\subsection{Perturbative Treatment}
In this subsection, we present two specific examples in which we compute perturbatively the expression for the reflected entropy and other correlation measures. These two examples correspond to $\beta=0$ and $\beta=\frac{\pi}{2}$ where due to the reflection symmetry, the profile of $\Sigma_{A\cup B}$ can be find exactly  at zero temperature. Using this result, we can evaluate the thermal corrections to reflected entropy at low temperature.

\subsection*{5.2.1 $\beta=0$}

In this case, the width of the entangling region lies along the anisotropic direction. In order to investigate the low temperature behavior of reflected entropy, we insert eq. \eqref{Liffunc} in eqs. \eqref{GS} and \eqref{GEW} and expand the resultant expressions in $hT\ll \ell T\ll 1$ limit which corresponds to $r_d\ll  r_u\ll \mu^{-\frac{3}{11}}$. Hence, the corresponding turning points are close to the boundary. It is straightforward to evaluate the leading order correction with the result
\begin{equation}
\ell=r_t\left( c+\frac{3 \sqrt{\pi } \,\Gamma \left(\frac{11}{8}\right) }{14 \,\Gamma \left(\frac{7}{8}\right)} \mu r_t^{11/3}\right),
\end{equation}
where $c=\frac{2 \sqrt{\pi } \Gamma \left(\frac{11}{16}\right)}{\Gamma \left(\frac{3}{16}\right)}>0$. Inverting the above equation, we can represent the turning point as a function of $\ell$
\begin{equation}\label{pertLifrtbeta0}
r_t=\frac{\ell}{c}\left(1-\frac{3 \sqrt{\pi } \,\Gamma \left(\frac{11}{8}\right) }{14 \,c^{14/3} \Gamma \left(\frac{7}{8}\right)}\mu\ell^{\frac{11}{3}}\right).
\end{equation}
That is, increasing the temperature, the turning point of the RT hypersurface decreases. In this limit, the leading order behavior of HEE reduces to
\begin{eqnarray}\label{heebb}
S=\frac{\tilde{R}^3L^2}{2G_N}\frac35 \left(\frac{1}{\epsilon^{\frac53}}-\frac{c}{2r_t^{\frac53}}+ \frac{5\sqrt{\pi}}{12}\frac{\Gamma\left(\frac{11}{8}\right)}{\Gamma\left(\frac{7}{8}\right)}\mu \,r_t^2\right).
\end{eqnarray}
Now we would like to recast this result in terms of boundary quantities. We do so by combining eqs. \eqref{pertLifrtbeta0} and \eqref{heebb} which allow us to translate the first order correction of HEE to the form
\begin{eqnarray}\label{svarBB}
\Delta S\equiv S-S_{\rm vac.}=\tilde{c}\,L^{2}\ell^2\varepsilon,
\end{eqnarray}
where $S_{\rm vac.}$ is the vacuum contribution given by $S_{\rm vac.}=\frac{3\tilde{R}^3L^2}{10G_N}\left(\frac{1}{\epsilon^{\frac53}}-\frac{c^{\frac83}}{2\ell^{\frac53}}\right)$ and $\tilde{c}=\frac{9\pi^{3/2}}{56 c^2}\frac{\Gamma\left(\frac{3}{8}\right)}{\Gamma\left(\frac{7}{8}\right)}$.
Note that $\tilde{c}>0$ and hence thermal excitations increase the HEE as expected. These results allow us to find the variation of HMI as follows
\begin{eqnarray}\label{dIadsBB}
\Delta I\equiv I-I_{\rm vac.}=- 2\tilde{c}L^{2}\left(\ell+h\right)^2\varepsilon,
\end{eqnarray}
where $I_{\rm vac.}$ is the vacuum contribution given by $I_{\rm vac.}=-\frac{3\tilde{R}^3L^2c^{\frac83}}{20G_N}\left(\frac{2}{\ell^{\frac53}}-\frac{1}{h^{\frac53}}-\frac{1}{(2\ell+h)^{\frac53}}\right)$. The minus sign shows that the thermal excitations decrease the HMI and hence reduce the total correlation between the subregions.
Finally, we turn to the thermal corrections to the reflected entropy. It is straightforward to carry out the perturbative analysis and we find that
\begin{eqnarray}
S_R=\frac{3\tilde{R}^3L^2}{10G_N}\left(\frac{1}{r_d^{\frac53}}-\frac{1}{r_u^{\frac53}}\right)+\frac{\tilde{R}^3L^2}{8G_N}\mu\left(r_u^2-r_d^2\right).
\end{eqnarray}
Now using eq. \eqref{pertLifrtbeta0} the leading contribution becomes
\begin{eqnarray}\label{ewvarBB}
\Delta {S_R}\equiv {S_R}-{S_R}_{\rm vac.}=-\mathcal{C}L^{2}\ell(\ell+h)\varepsilon,
\end{eqnarray}
where ${S_R}_{\rm vac.}$ is the vacuum contribution given by ${S_R}_{\rm vac.}=\frac{3\tilde{R}^3L^2c^{\frac53}}{10G_N}\left(\frac{1}{h^{\frac53}}-\frac{1}{(2\ell+h)^{\frac53}}\right)$ and $\mathcal{C}=\frac{3\pi}{c^2}\left(\frac{6\sqrt{\pi}}{7c}\frac{\Gamma\left(\frac{11}{8}\right)}{\Gamma\left(\frac{7}{8}\right)}-1\right)$. We note again that this contribution is negative and hence the finite temperature corrections decrease the reflected entropy. Regarding this quantity as a measure of total correlation between the two subregions, we see that thermal excitations promote disentangling between them. These results are consistent with the previously numerical results illustrated in figure \ref{fig:LifMIEW}.

\subsection*{5.2.2 $\beta=\pi/2$}
The analysis follows similarly to the previous case, with the obvious replacement of the rotation angle. Hence we just report the final results in what follows. At leading order the variation of HEE becomes
\begin{equation*}
\Delta S\equiv S-S^{\pi/2}_{\rm vac}=L^2\tilde{c}  \,\ell^{\frac{5}{2}}\varepsilon,
\end{equation*}
where $\tilde{c}= \frac{36 \pi^{3/2}\Gamma \left(\frac{21}{16}\right) }{65 \Gamma \left(\frac{13}{16}\right)}\left(\frac{2}{3c}\right)^{5/2}$, $c=\frac{2 \sqrt{\pi } \Gamma \left(\frac{5}{8}\right)}{\Gamma \left(\frac{1}{8}\right)}$  and $S^{\pi/2}_{\rm vac}$ is the vacuum contribution in $\beta=\frac{\pi}{2}$ given by
\begin{equation}
S^{\pi/2}_{\rm vac}=\frac{\tilde{R}^3L^2}{8G_N}\left(\frac{2}{ {\epsilon}^2}-\frac{c^4}{(\frac{2\ell}{3})^3}\right).
\end{equation}
Equipped with the above result we can compute HMI as follows
\begin{eqnarray}\label{dIadsBB1}
\Delta I\equiv I-I^{\pi/2}_{\rm vac}=-L^2\tilde{c}\left((2\ell+h)^{5/2}-2\ell^{5/2}+h^{5/2}\right)\varepsilon,
\end{eqnarray}
where $I^{\pi/2}_{\rm vac}$ is the vacuum contribution in $\beta=\frac{\pi}{2}$ given by
\begin{equation}
I^{\pi/2}_{\rm vac}=-\frac{\tilde{R}^3L^2}{12 G}\left(\frac{3\sqrt{\pi}\Gamma\left(\frac{5}{8}\right)}{\Gamma\left(\frac{1}{8}\right)}\right)^4\left(\frac{2}{\ell^{3}}-\frac{1}{h^{3}}-\frac{1}{(2\ell+h)^{3}}\right).
\end{equation}
Finally, the variation of reflected entropy becomes
\begin{eqnarray}\label{ewvarBB1}
\Delta S_R\equiv S_R-{S_R}^{\pi/2}_{\rm vac}=-\tilde{\mathcal{C}}L^2\left((h+2 l)^{5/2}-h^{5/2}\right)\varepsilon,
\end{eqnarray}
where
$\tilde{\mathcal{C}}=\frac{9\pi}{10 c^{5/2}}\left(\frac{2}{3}\right)^{5/2}\left(\frac{10 \sqrt{\pi } \Gamma \left(\frac{21}{16}\right)}{13 c \Gamma \left(\frac{13}{16}\right)}-1\right)$ and ${S_R}^{\pi/2}_{\rm vac}=\frac{27\tilde{R}^3 L^2}{32G_N}c^{3}\left(\frac{1}{h^{3}}-\frac{1}{(2\ell+h)^{3}}\right)$. Again, these agree with the results shown
in figure \ref{fig:LifMIEW}, where we see that thermal excitations promote disentangling between the subregions.   

%Let us mention that a remarkable feature of this gravitational background is the appearance of a universal logarithmic term for related to the holographic entanglement entropy divergent part. Unlike parallel and orthogonal cases, in other angles the divergent term includes a .

\section{Conclusions and Discussions}\label{sec:diss}

In this paper, we explored the behavior of reflected entropy in certain nonrelativistic geometries dual to anisotropic boundary systems. We used the holographic proposal for computing this quantity which states that reflected entropy is proportional to the minimal cross-sectional area of the entanglement wedge,
as in eq. \eqref{HRE}. Specifically, we have focused on symmetric boundary configurations consisting of two disjoint strips with equal width, which is the simplest case to utilize the holographic proposal to compute the correlation measures. In principle though, we expect that the qualitative
features of our results are independent of the specific configuration. Although some of the intermediate steps may differ, we expect that the qualitative features of our results are hold for non symmetric configurations. In addition to numerical analysis, in specific anisotropic backgrounds we evaluated the leading order corrections to holographic correlation measures analytically. We also compared the behavior of reflected entropy to other correlation measures including HEE and HMI.

Our analysis in this paper focused mainly on the effect of anisotropy on reflected entropy and other correlation measures. Generally, the additional contributions due to the anisotropy parameter or rotation angle to this quantity do not have a definite sign. For example, based on our results in section \ref{sec:condecon}, we found that $S_R$ is an increasing function of the anisotropy parameter, \textit{i.e.,} $\nu$, in a specific range of rotation angle  such that it develops a maximum at $\beta=\pi/2$. Interestingly, at the same value of the rotation angle, the critical separation between the subregions is a monotonically increasing function of $\nu$ and hence the correlation between the subregions becomes stronger (see figure \ref{fig:numfig3}). On the other hand, both our analytic calculations and numerical analysis in section \ref{sec:nonconformal} gave evidence that the reflected entropy has a minimum at $\beta=\pi/2$ and is a monotonically decreasing function of the anisotropy parameter (see the panel in figure \ref{fig:EADS} and eq. \eqref{srpertnonconformal}).

In addition to these differences, at a qualitative level, all of the cases considered in this work had a number of common features in all cases examples. First, the variation of HMI and reflected entropy has the sign due to presence of the anisotropy. Regarding these quantities as measures of total correlation between the subregions,
this behavior seems reasonable. Although, this result is different from what happens for the HEE
where the variation flips its sign. This feature precisely matches with the previous results of \cite{Sahraei:2021wqn,Blanco:2013joa,Alishahiha:2014jxa}. Another key feature which was observed here was the appearance of a new universal logarithmic term in HEE whose coefficient depends on the anisotropy parameter and the rotation angle. Roughly, we can think of this universal term as characterizing when the isotropy is broken in the underlying boundary theory. Similarly, as shown in [55], if instead we choose a background which breaks the
translation invariance the structure of the universal terms of HEE is modified. This feature is entirely expected given our experiences from HEE in other backgrounds with broken symmetries, \textit{e.g.,} see \cite{RezaMohammadiMozaffar:2016lbo}.

\begin{figure}[h]
\begin{center}
\includegraphics[scale=0.6]{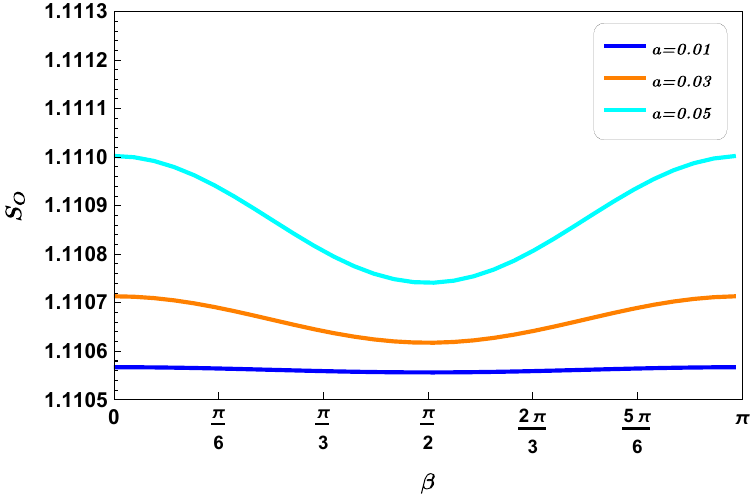}
\hspace*{1cm}
\includegraphics[scale=0.6]{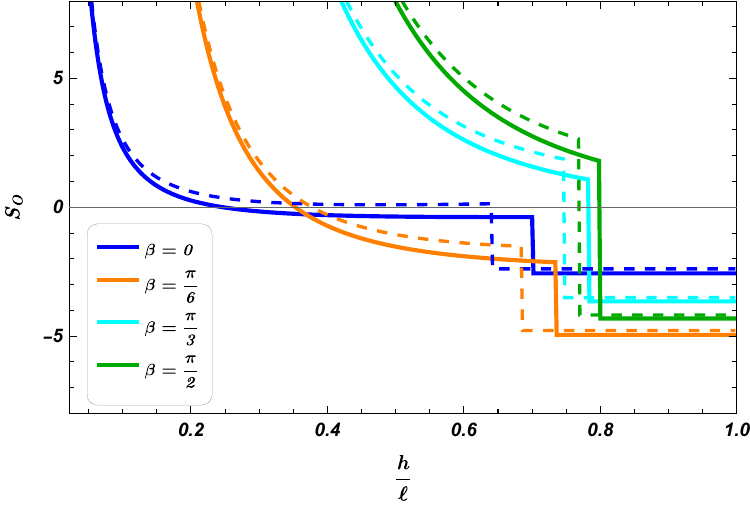}
\end{center}
\caption{ \textit{Left}: Odd entropy as a function of $\beta$ for different values of the anisotropy parameter for model discussed in section \ref{sec:nonconformal}. \textit{Right}: Odd entropy as a function of $\frac{h}{\ell}$ for different values of $\beta$ for model discussed in section \ref{sec:Lifshitz}. The dashed curves represent the finite temperature results and the solid curves correspond to $T=0$ case.}
\label{fig:odd}
\end{figure}

Recall that some holographic proposals consider other candidates for the mixed state correlation measures dual to EWCS. For example, based on \cite{Tamaoka:2018ned}, in holographic theories the odd entropy can be written in terms of the reflected entropy as follows
\begin{eqnarray*}
S_O(A, B)=\frac{S_R(A, B)}{2}+S(A\cup B).
\end{eqnarray*}
Using this expression we can find the odd entropy using our previous results for reflected entropy in different anisotropic boundary systems. In figure \ref{fig:odd} we plot this measure in two of our models in sections  
\ref{sec:nonconformal} and \ref{sec:Lifshitz}. The left panel shows that in the non-conformal boundary theory, odd entropy is an increasing function of the anisotropy parameter and hence the correlation between the subregions becomes more pronounced. Clearly the $\beta$ dependence of this measure is similar to HMI and reflected entropy as expected (see figure \ref{fig:EADS}). In the right panel we plot odd entropy as a function of $\frac{h}{\ell}$ for different values of $\beta$ in the Lifshitz-like anisotropic background. The dashed curves represent the finite temperature results and
the solid curves correspond to the zero temperature case. Regarding the odd entropy as a measure of correlation, we see that decreasing the rotation angle promote disentangling between the subregions. Let us mention that the qualitative features of the odd entropy in other anisotropic models are similar, thus we neglect to present them.
%Of course, one can compute the odd entropy in other cases, but in these examples we also have an analytic expression for this quantity.  

We can extend this study to different interesting directions. In this paper we focused on symmetric configuration for the boundary
entangling regions which significantly simplifies the computation of the reflected entropy. It is also interesting to look at more complicated setups where the widths of the strips are different, using the method
first introduced in \cite{Liu:2019qje}. Another interesting question is if either of these behaviors can be extracted from field theory calculations of reflected entropy using the techniques developed in \cite{Bueno:2020fle,Camargo:2021aiq}. We plan to explore some of these
directions in the near future.

\subsection*{Acknowledgements}
We are very grateful to Mohammad Hasan Vahidinia for careful reading
of the manuscript and his valuable comments.


\begin{thebibliography}{}
%\cite{Ryu:2006bv}
\bibitem{Ryu:2006bv}
S.~Ryu and T.~Takayanagi,
``Holographic derivation of entanglement entropy from AdS/CFT,''
Phys. Rev. Lett. \textbf{96}, 181602 (2006)
doi:10.1103/PhysRevLett.96.181602
[arXiv:hep-th/0603001 [hep-th]].
%2900 citations counted in INSPIRE as of 09 Jan 2022

%\cite{Brown:2015bva}
\bibitem{Brown:2015bva}
A.~R.~Brown, D.~A.~Roberts, L.~Susskind, B.~Swingle and Y.~Zhao,
``Holographic Complexity Equals Bulk Action?,''
Phys. Rev. Lett. \textbf{116}, no.19, 191301 (2016)
doi:10.1103/PhysRevLett.116.191301
[arXiv:1509.07876 [hep-th]].
%544 citations counted in INSPIRE as of 10 Jan 2022

%\cite{Casini:2009sr}
\bibitem{Casini:2009sr}
H.~Casini and M.~Huerta,
``Entanglement entropy in free quantum field theory,''
J. Phys. A \textbf{42}, 504007 (2009)
doi:10.1088/1751-8113/42/50/504007
[arXiv:0905.2562 [hep-th]].
%501 citations counted in INSPIRE as of 15 Jan 2022

%\cite{Nishioka:2018khk}
\bibitem{Nishioka:2018khk}
T.~Nishioka,
``Entanglement entropy: holography and renormalization group,''
Rev. Mod. Phys. \textbf{90}, no.3, 035007 (2018)
doi:10.1103/RevModPhys.90.035007
[arXiv:1801.10352 [hep-th]].
%120 citations counted in INSPIRE as of 09 Jan 2022



%\cite{Nishioka:2009un}
\bibitem{Nishioka:2009un}
T.~Nishioka, S.~Ryu and T.~Takayanagi,
``Holographic Entanglement Entropy: An Overview,''
J. Phys. A \textbf{42}, 504008 (2009)
doi:10.1088/1751-8113/42/50/504008
[arXiv:0905.0932 [hep-th]].
%634 citations counted in INSPIRE as of 10 Jan 2022


%\cite{Rangamani:2016dms}
\bibitem{Rangamani:2016dms}
M.~Rangamani and T.~Takayanagi,
``Holographic Entanglement Entropy,''
Lect. Notes Phys. \textbf{931}, pp.1-246 (2017)
doi:10.1007/978-3-319-52573-0
[arXiv:1609.01287 [hep-th]].
%219 citations counted in INSPIRE as of 09 Jan 2022

%\cite{Plenio:2005cwa}
\bibitem{Plenio:2005cwa}
M.~B.~Plenio,
``Logarithmic Negativity: A Full Entanglement Monotone That is not Convex,''
Phys. Rev. Lett. \textbf{95}, no.9, 090503 (2005)
doi:10.1103/PhysRevLett.95.090503
[arXiv:quant-ph/0505071 [quant-ph]].
%245 citations counted in INSPIRE as of 16 Jan 2022

%\cite{Terhal:2002}
\bibitem{Terhal:2002}
B. M. Terhal, M. Horodecki, D. W. Leung and D. P. Di-
Vincenzo, The entanglement of purification," J. Math.
Phys. 43 (2002) 4286, arXiv:quant-ph/0202044.

%\cite{Tamaoka:2018ned,BabaeiVelni:2019pkw,Jokela:2019ebz,Dutta:2019gen,Amrahi:2020jqg,BabaeiVelni:2020wfl}
\bibitem{Tamaoka:2018ned}
K.~Tamaoka,
``Entanglement Wedge Cross Section from the Dual Density Matrix,''
Phys. Rev. Lett. \textbf{122}, no.14, 141601 (2019)
doi:10.1103/PhysRevLett.122.141601
[arXiv:1809.09109 [hep-th]].
%66 citations counted in INSPIRE as of 08 Feb 2021

%\cite{Dutta:2019gen}
\bibitem{Dutta:2019gen}
S.~Dutta and T.~Faulkner,
``A canonical purification for the entanglement wedge cross-section,''
JHEP \textbf{03}, 178 (2021)
doi:10.1007/JHEP03(2021)178
[arXiv:1905.00577 [hep-th]].
%85 citations counted in INSPIRE as of 21 May 2021


%\cite{Takayanagi:2017knl,Nguyen:2017yqw,Kusuki:2019zsp,Tamaoka:2018ned}
\bibitem{Takayanagi:2017knl}
T.~Takayanagi and K.~Umemoto,
``Entanglement of purification through holographic duality,''
Nature Phys. \textbf{14}, no.6, 573-577 (2018)
doi:10.1038/s41567-018-0075-2
[arXiv:1708.09393 [hep-th]].
%149 citations counted in INSPIRE as of 08 Feb 2021

%\cite{Nguyen:2017yqw}
\bibitem{Nguyen:2017yqw}
P.~Nguyen, T.~Devakul, M.~G.~Halbasch, M.~P.~Zaletel and B.~Swingle,
``Entanglement of purification: from spin chains to holography,''
JHEP \textbf{01}, 098 (2018)
doi:10.1007/JHEP01(2018)098
[arXiv:1709.07424 [hep-th]].
%113 citations counted in INSPIRE as of 08 Feb 2021


%\cite{Kusuki:2019zsp}
\bibitem{Kusuki:2019zsp}
Y.~Kusuki, J.~Kudler-Flam and S.~Ryu,
``Derivation of Holographic Negativity in AdS$_3$/CFT$_2$,''
Phys. Rev. Lett. \textbf{123}, no.13, 131603 (2019)
doi:10.1103/PhysRevLett.123.131603
[arXiv:1907.07824 [hep-th]].
%34 citations counted in INSPIRE as of 08 Feb 2021






%\cite{Hirai:2018jwy}
\bibitem{Hirai:2018jwy}
H.~Hirai, K.~Tamaoka and T.~Yokoya,
``Towards Entanglement of Purification for Conformal Field Theories,''
PTEP \textbf{2018}, no.6, 063B03 (2018)
doi:10.1093/ptep/pty063
[arXiv:1803.10539 [hep-th]].
%46 citations counted in INSPIRE as of 25 May 2021

%\cite{BabaeiVelni:2019pkw,Jokela:2019ebz,Dutta:2019gen,Amrahi:2020jqg,BabaeiVelni:2020wfl}
\bibitem{BabaeiVelni:2019pkw} 
K.~Babaei Velni, M.~R.~Mohammadi Mozaffar and M.~H.~Vahidinia,
``Some Aspects of Entanglement Wedge Cross-Section,''
JHEP {\bf 1905}, 200 (2019)
doi:10.1007/JHEP05(2019)200
[arXiv:1903.08490 [hep-th]].
%%CITATION = doi:10.1007/JHEP05(2019)200;%%
%16 citations counted in INSPIRE as of 01 May 2020



%\cite{Jokela:2019ebz,Dutta:2019gen,Amrahi:2020jqg,BabaeiVelni:2020wfl}
\bibitem{Jokela:2019ebz}
N.~Jokela and A.~P\"onni,
``Notes on entanglement wedge cross sections,''
JHEP \textbf{07}, 087 (2019)
doi:10.1007/JHEP07(2019)087
[arXiv:1904.09582 [hep-th]].
%27 citations counted in INSPIRE as of 08 Feb 2021

%\cite{Umemoto:2019jlz}
\bibitem{Umemoto:2019jlz}
K.~Umemoto,
``Quantum and Classical Correlations Inside the Entanglement Wedge,''
Phys. Rev. D \textbf{100}, no.12, 126021 (2019)
doi:10.1103/PhysRevD.100.126021
[arXiv:1907.12555 [hep-th]].
%25 citations counted in INSPIRE as of 23 May 2021


%\cite{Akers:2019gcv}
\bibitem{Akers:2019gcv}
C.~Akers and P.~Rath,
``Entanglement Wedge Cross Sections Require Tripartite Entanglement,''
JHEP \textbf{04}, 208 (2020)
doi:10.1007/JHEP04(2020)208
[arXiv:1911.07852 [hep-th]].
%26 citations counted in INSPIRE as of 23 May 2021

%\cite{Amrahi:2020jqg,BabaeiVelni:2020wfl}
\bibitem{Amrahi:2020jqg}
B.~Amrahi, M.~Ali-Akbari and M.~Asadi,
``Holographic entanglement of purification near a critical point,''
Eur. Phys. J. C \textbf{80}, no.12, 1152 (2020)
doi:10.1140/epjc/s10052-020-08647-8
[arXiv:2004.02856 [hep-th]].
%4 citations counted in INSPIRE as of 08 Feb 2021

%\cite{Chakrabortty:2020ptb}
\bibitem{Chakrabortty:2020ptb}
S.~Chakrabortty, S.~Pant and K.~Sil,
``Effect of back reaction on entanglement and subregion volume complexity in strongly coupled plasma,''
JHEP \textbf{06}, 061 (2020)
doi:10.1007/JHEP06(2020)061
[arXiv:2004.06991 [hep-th]].
%8 citations counted in INSPIRE as of 31 May 2021

%\cite{Saha:2021kwq}
\bibitem{Saha:2021kwq}
A.~Saha and S.~Gangopadhyay,
``Holographic study of entanglement and complexity for mixed states,''
Phys. Rev. D \textbf{103}, no.8, 086002 (2021)
doi:10.1103/PhysRevD.103.086002
[arXiv:2101.00887 [hep-th]].
%1 citations counted in INSPIRE as of 31 May 2021

%\cite{1810.00420}
\bibitem{1810.00420} 
R.~Q.~Yang, C.~Y.~Zhang and W.~M.~Li,
``Holographic entanglement of purification for thermofield double states and thermal quench,''
JHEP {\bf 1901}, 114 (2019)
doi:10.1007/JHEP01(2019)114
[arXiv:1810.00420 [hep-th]].
%%CITATION = doi:10.1007/JHEP01(2019)114;%%
%21 citations counted in INSPIRE as of 01 May 2020



%\cite{1907.06646}
\bibitem{1907.06646} 
Y.~Kusuki and K.~Tamaoka,
``Dynamics of Entanglement Wedge Cross Section from Conformal Field Theories,''
arXiv:1907.06646 [hep-th].
%%CITATION = ARXIV:1907.06646;%%
%16 citations counted in INSPIRE as of 01 May 2020



%\cite{Jeong:2019xdr}
\bibitem{Jeong:2019xdr}
H.~S.~Jeong, K.~Y.~Kim and M.~Nishida,
``Reflected Entropy and Entanglement Wedge Cross Section with the First Order Correction,''
JHEP \textbf{12}, 170 (2019)
doi:10.1007/JHEP12(2019)170
[arXiv:1909.02806 [hep-th]].
%32 citations counted in INSPIRE as of 09 Aug 2022


%\cite{2001.05501}
\bibitem{2001.05501} 
J.~Kudler-Flam, Y.~Kusuki and S.~Ryu,
``Correlation measures and the entanglement wedge cross-section after quantum quenches in two-dimensional conformal field theories,''
JHEP {\bf 2004}, 074 (2020)
doi:10.1007/JHEP04(2020)074
[arXiv:2001.05501 [hep-th]].
%%CITATION = doi:10.1007/JHEP04(2020)074;%%
%4 citations counted in INSPIRE as of 01 May 2020



%\cite{BabaeiVelni:2020wfl}
\bibitem{BabaeiVelni:2020wfl}
K.~Babaei Velni, M.~R.~Mohammadi Mozaffar and M.~H.~Vahidinia,
``Evolution of entanglement wedge cross section following a global quench,''
JHEP \textbf{08}, 129 (2020)
doi:10.1007/JHEP08(2020)129
[arXiv:2005.05673 [hep-th]].
%7 citations counted in INSPIRE as of 08 Feb 2021



%\cite{Moosa:2020vcs,Kudler-Flam:2020url}
\bibitem{Moosa:2020vcs}
M.~Moosa,
``Time dependence of reflected entropy in conformal field theory,''
[arXiv:2001.05969 [hep-th]].
%2 citations counted in INSPIRE as of 07 May 2020

%\cite{Boruch:2020wbe}
\bibitem{Boruch:2020wbe}
J.~Boruch,
``Entanglement wedge cross-section in shock wave geometries,''
JHEP \textbf{07}, 208 (2020)
doi:10.1007/JHEP07(2020)208
[arXiv:2006.10625 [hep-th]].
%4 citations counted in INSPIRE as of 21 May 2021

%\cite{Basu:2022nds,Basak:2022cjs}
\bibitem{Basu:2022nds}
D.~Basu, H.~Parihar, V.~Raj and G.~Sengupta,
``Entanglement negativity, reflected entropy, and anomalous gravitation,''
Phys. Rev. D \textbf{105}, no.8, 086013 (2022)
[erratum: Phys. Rev. D \textbf{105}, no.12, 129902 (2022)]
doi:10.1103/PhysRevD.105.086013
[arXiv:2202.00683 [hep-th]].
%9 citations counted in INSPIRE as of 09 Aug 2022

%\cite{Basak:2022cjs}
\bibitem{Basak:2022cjs}
J.~K.~Basak, H.~Chourasiya, V.~Raj and G.~Sengupta,
``Reflected entropy in Galilean conformal field theories and flat holography,''
[arXiv:2202.01201 [hep-th]].
%5 citations counted in INSPIRE as of 09 Aug 2022



%\cite{1909.06790}
\bibitem{1909.06790}
Y.~Kusuki and K.~Tamaoka,
``Entanglement Wedge Cross Section from CFT: Dynamics of Local Operator Quench,''
JHEP \textbf{02}, 017 (2020)
doi:10.1007/JHEP02(2020)017
[arXiv:1909.06790 [hep-th]].
%12 citations counted in INSPIRE as of 07 May 2020



%\cite{Mollabashi:2020ifv,:2020fle,Camargo:2020yfv,Camargo:2021aiq}
\bibitem{Mollabashi:2020ifv}
A.~Mollabashi and K.~Tamaoka,
``A Field Theory Study of Entanglement Wedge Cross Section: Odd Entropy,''
JHEP \textbf{08}, 078 (2020)
doi:10.1007/JHEP08(2020)078
[arXiv:2004.04163 [hep-th]].
%4 citations counted in INSPIRE as of 08 Feb 2021

%\cite{Berthiere:2020ihq}
\bibitem{Berthiere:2020ihq}
C.~Berthiere, H.~Chen, Y.~Liu and B.~Chen,
``Topological reflected entropy in Chern-Simons theories,''
Phys. Rev. B \textbf{103}, no.3, 035149 (2021)
doi:10.1103/PhysRevB.103.035149
[arXiv:2008.07950 [hep-th]].
%8 citations counted in INSPIRE as of 09 Aug 2022

%\cite{Bueno:2020fle,Camargo:2020yfv,Camargo:2021aiq}
\bibitem{Bueno:2020fle}
P.~Bueno and H.~Casini,
``Reflected entropy for free scalars,''
JHEP \textbf{11}, 148 (2020)
doi:10.1007/JHEP11(2020)148
[arXiv:2008.11373 [hep-th]].
%0 citations counted in INSPIRE as of 08 Feb 2021


%\cite{Camargo:2021aiq}
\bibitem{Camargo:2021aiq}
H.~A.~Camargo, L.~Hackl, M.~P.~Heller, \textdaggerdbl{}.~A.~Jahn and B.~Windt,
``Long-distance entanglement of purification and reflected entropy in conformal field theory,''
[arXiv:2102.00013 [hep-th]].
%0 citations counted in INSPIRE as of 21 May 2021


%\cite{Wen:2021qgx}
\bibitem{Wen:2021qgx}
Q.~Wen,
``Balanced Partial Entanglement and the Entanglement Wedge Cross Section,''
JHEP \textbf{04}, 301 (2021)
doi:10.1007/JHEP04(2021)301
[arXiv:2103.00415 [hep-th]].
%2 citations counted in INSPIRE as of 21 May 2021



%\cite{Hayden:2021gno}
\bibitem{Hayden:2021gno}
P.~Hayden, O.~Parrikar and J.~Sorce,
``The Markov gap for geometric reflected entropy,''
JHEP \textbf{10}, 047 (2021)
doi:10.1007/JHEP10(2021)047
[arXiv:2107.00009 [hep-th]].
%6 citations counted in INSPIRE as of 16 Jan 2022



%\cite{Akers:2021pvd}
\bibitem{Akers:2021pvd}
C.~Akers, T.~Faulkner, S.~Lin and P.~Rath,
``Reflected entropy in random tensor networks,''
[arXiv:2112.09122 [hep-th]].
%0 citations counted in INSPIRE as of 16 Jan 2022

%\cite{Bueno:2020vnx}
\bibitem{Bueno:2020vnx}
P.~Bueno and H.~Casini,
``Reflected entropy, symmetries and free fermions,''
JHEP \textbf{05}, 103 (2020)
doi:10.1007/JHEP05(2020)103
[arXiv:2003.09546 [hep-th]].
%14 citations counted in INSPIRE as of 16 Jan 2022



%\cite{Camargo:2022mme}
\bibitem{Camargo:2022mme}
H.~A.~Camargo, P.~Nandy, Q.~Wen and H.~Zhong,
``Balanced partial entanglement and mixed state correlations,''
SciPost Phys. \textbf{12}, no.4, 137 (2022)
doi:10.21468/SciPostPhys.12.4.137
[arXiv:2201.13362 [hep-th]].
%6 citations counted in INSPIRE as of 09 Aug 2022

%\cite{Mateos:2011tv,Rougemont:2015oea,Rebhan:2011vd,Mateos:2011ix,Giataganas:2012zy,Chernicoff:2012iq,Ali-Akbari:2013txa}
\bibitem{Mateos:2011tv}
D.~Mateos and D.~Trancanelli,
``Thermodynamics and Instabilities of a Strongly Coupled Anisotropic Plasma,''
JHEP \textbf{07}, 054 (2011)
doi:10.1007/JHEP07(2011)054
[arXiv:1106.1637 [hep-th]].
%177 citations counted in INSPIRE as of 18 Jan 2022

%\cite{Rougemont:2015oea,Rebhan:2011vd,Mateos:2011ix,Giataganas:2012zy,Chernicoff:2012iq,Ali-Akbari:2013txa}
\bibitem{Rougemont:2015oea}
R.~Rougemont, R.~Critelli and J.~Noronha,
``Holographic calculation of the QCD crossover temperature in a magnetic field,''
Phys. Rev. D \textbf{93}, no.4, 045013 (2016)
doi:10.1103/PhysRevD.93.045013
[arXiv:1505.07894 [hep-th]].
%79 citations counted in INSPIRE as of 18 Jan 2022

%\cite{Rebhan:2011vd,Mateos:2011ix,Giataganas:2012zy,Chernicoff:2012iq,Ali-Akbari:2013txa}
\bibitem{Rebhan:2011vd}
A.~Rebhan and D.~Steineder,
``Violation of the Holographic Viscosity Bound in a Strongly Coupled Anisotropic Plasma,''
Phys. Rev. Lett. \textbf{108}, 021601 (2012)
doi:10.1103/PhysRevLett.108.021601
[arXiv:1110.6825 [hep-th]].
%193 citations counted in INSPIRE as of 18 Jan 2022


%\cite{Mateos:2011ix,Giataganas:2012zy,Chernicoff:2012iq,Ali-Akbari:2013txa}
\bibitem{Mateos:2011ix}
D.~Mateos and D.~Trancanelli,
``The anisotropic N=4 super Yang-Mills plasma and its instabilities,''
Phys. Rev. Lett. \textbf{107}, 101601 (2011)
doi:10.1103/PhysRevLett.107.101601
[arXiv:1105.3472 [hep-th]].
%161 citations counted in INSPIRE as of 18 Jan 2022


%\cite{Giataganas:2012zy,Chernicoff:2012iq,Ali-Akbari:2013txa}
\bibitem{Giataganas:2012zy}
D.~Giataganas,
``Probing strongly coupled anisotropic plasma,''
JHEP \textbf{07}, 031 (2012)
doi:10.1007/JHEP07(2012)031
[arXiv:1202.4436 [hep-th]].
%129 citations counted in INSPIRE as of 18 Jan 2022


%\cite{Chernicoff:2012iq,Ali-Akbari:2013txa}
\bibitem{Chernicoff:2012iq}
M.~Chernicoff, D.~Fernandez, D.~Mateos and D.~Trancanelli,
``Drag force in a strongly coupled anisotropic plasma,''
JHEP \textbf{08}, 100 (2012)
doi:10.1007/JHEP08(2012)100
[arXiv:1202.3696 [hep-th]].
%74 citations counted in INSPIRE as of 18 Jan 2022



%\cite{Ali-Akbari:2013txa}
\bibitem{Ali-Akbari:2013txa}
M.~Ali-Akbari and H.~Ebrahim,
``Chiral symmetry breaking: To probe anisotropy and magnetic field in quark-gluon plasma,''
Phys. Rev. D \textbf{89}, no.6, 065029 (2014)
doi:10.1103/PhysRevD.89.065029
[arXiv:1309.4715 [hep-th]].
%22 citations counted in INSPIRE as of 18 Jan 2022


%\cite{Pal:2009yp,Azeyanagi:2009pr}
\bibitem{Pal:2009yp}
S.~S.~Pal,
``Anisotropic gravity solutions in AdS/CMT,''
[arXiv:0901.0599 [hep-th]].
%32 citations counted in INSPIRE as of 24 Jul 2022

%\cite{Azeyanagi:2009pr}
\bibitem{Azeyanagi:2009pr}
T.~Azeyanagi, W.~Li and T.~Takayanagi,
``On String Theory Duals of Lifshitz-like Fixed Points,''
JHEP \textbf{06}, 084 (2009)
doi:10.1088/1126-6708/2009/06/084
[arXiv:0905.0688 [hep-th]].
%171 citations counted in INSPIRE as of 24 Jul 2022
%\cite{Narayan:2012ks,Jahnke:2017iwi,Sahraei:2021wqn}
\bibitem{Narayan:2012ks}
K.~Narayan, T.~Takayanagi and S.~P.~Trivedi,
``AdS plane waves and entanglement entropy,''
JHEP \textbf{04}, 051 (2013)
doi:10.1007/JHEP04(2013)051
[arXiv:1212.4328 [hep-th]].
%39 citations counted in INSPIRE as of 07 Mar 2021
%\cite{Jahnke:2017iwi,Sahraei:2021wqn}
\bibitem{Jahnke:2017iwi}
V.~Jahnke,
``Delocalizing entanglement of anisotropic black branes,''
JHEP \textbf{01}, 102 (2018)
doi:10.1007/JHEP01(2018)102
[arXiv:1708.07243 [hep-th]].
%21 citations counted in INSPIRE as of 18 Jan 2022

%\cite{Mahapatra:2019uql}
\bibitem{Mahapatra:2019uql}
S.~Mahapatra,
``Interplay between the holographic QCD phase diagram and mutual \textbackslash{}\& $n$-partite information,''
JHEP \textbf{04}, 137 (2019)
doi:10.1007/JHEP04(2019)137
[arXiv:1903.05927 [hep-th]].
%19 citations counted in INSPIRE as of 09 Aug 2022


%\cite{Arefeva:2020uec}
\bibitem{Arefeva:2020uec}
I.~Y.~Aref'eva, A.~Patrushev and P.~Slepov,
``Holographic entanglement entropy in anisotropic background with confinement-deconfinement phase transition,''
JHEP \textbf{07}, 043 (2020)
doi:10.1007/JHEP07(2020)043
[arXiv:2003.05847 [hep-th]].
%27 citations counted in INSPIRE as of 28 Jul 2022

%\cite{Jain:2020rbb}
\bibitem{Jain:2020rbb}
P.~Jain and S.~Mahapatra,
``Mixed state entanglement measures as probe for confinement,''
Phys. Rev. D \textbf{102}, 126022 (2020)
doi:10.1103/PhysRevD.102.126022
[arXiv:2010.07702 [hep-th]].
%14 citations counted in INSPIRE as of 09 Aug 2022


%\cite{Sahraei:2021wqn}
\bibitem{Sahraei:2021wqn}
M.~Sahraei, M.~J.~Vasli, M.~R.~M.~Mozaffar and K.~B.~Velni,
``Entanglement Wedge Cross Section in Holographic Excited States,''
doi:10.1007/JHEP08(2021)038
[arXiv:2105.12476 [hep-th]].
%4 citations counted in INSPIRE as of 18 Jan 2022
%\cite{Aref2018}
\bibitem{Aref2018}
I.~Aref'eva and K.~Rannu,
``Holographic Anisotropic Background with Confinement-Deconfinement Phase Transition,''
JHEP \textbf{05}, 206 (2018)
doi:10.1007/JHEP05(2018)206
[arXiv:1802.05652 [hep-th]].
%57 citations counted in INSPIRE as of 28 Jul 2022	
	%\cite{Gubser2008}
\bibitem{Gubser2008}
S.~S.~Gubser and A.~Nellore,
``Mimicking the QCD equation of state with a dual black hole,''
Phys. Rev. D \textbf{78}, 086007 (2008)
doi:10.1103/PhysRevD.78.086007
[arXiv:0804.0434 [hep-th]].
%256 citations counted in INSPIRE as of 28 Jul 2022	
%\cite{Nellore2008}
\bibitem{Nellore2008}
S.~S.~Gubser, A.~Nellore, S.~S.~Pufu and F.~D.~Rocha,
``Thermodynamics and bulk viscosity of approximate black hole duals to finite temperature quantum chromodynamics,''
Phys. Rev. Lett. \textbf{101}, 131601 (2008)
doi:10.1103/PhysRevLett.101.131601
[arXiv:0804.1950 [hep-th]].
%231 citations counted in INSPIRE as of 28 Jul 2022
%\cite{Giataganas2018}
\bibitem{Giataganas2018}
D.~Giataganas, U.~G\"ursoy and J.~F.~Pedraza,
``Strongly-coupled anisotropic gauge theories and holography,''
Phys. Rev. Lett. \textbf{121}, no.12, 121601 (2018)
doi:10.1103/PhysRevLett.121.121601
[arXiv:1708.05691 [hep-th]].
%82 citations counted in INSPIRE as of 28 Jul 2022	
	%\cite{Gursoy:2007er}
\bibitem{Gursoy:2007er}
U.~Gursoy, E.~Kiritsis and F.~Nitti,
``Exploring improved holographic theories for QCD: Part II,''
JHEP \textbf{02}, 019 (2008)
doi:10.1088/1126-6708/2008/02/019
[arXiv:0707.1349 [hep-th]].
%441 citations counted in INSPIRE as of 19 Jun 2022
 

%\cite{Ghasemi2019}
\bibitem{Ghasemi2019}
M.~Ghasemi and S.~Parvizi,
``Constraints on anisotropic RG flows from holographic entanglement entropy,''
[arXiv:1907.01546 [hep-th]].
%7 citations counted in INSPIRE as of 28 Jul 2022




%\cite{RezaMohammadiMozaffar:2016lbo}
\bibitem{RezaMohammadiMozaffar:2016lbo}
M.~Reza Mohammadi Mozaffar, A.~Mollabashi and F.~Omidi,
``Non-local Probes in Holographic Theories with Momentum Relaxation,''
JHEP \textbf{10}, 135 (2016)
doi:10.1007/JHEP10(2016)135
[arXiv:1608.08781 [hep-th]].
%18 citations counted in INSPIRE as of 18 Jul 2022




%\cite{Azeyanagi2009}
\bibitem{Azeyanagi2009}
T.~Azeyanagi, W.~Li and T.~Takayanagi,
``On String Theory Duals of Lifshitz-like Fixed Points,''
JHEP \textbf{06}, 084 (2009)
doi:10.1088/1126-6708/2009/06/084
[arXiv:0905.0688 [hep-th]].
%171 citations counted in INSPIRE as of 28 Jul 2022
	
	
	
		%\cite{Blanco:2013joa}
\bibitem{Blanco:2013joa}
D.~D.~Blanco, H.~Casini, L.~Y.~Hung and R.~C.~Myers,
``Relative Entropy and Holography,''
JHEP \textbf{08}, 060 (2013)
doi:10.1007/JHEP08(2013)060
[arXiv:1305.3182 [hep-th]].
%327 citations counted in INSPIRE as of 27 Jul 2022
	
	%\cite{Alishahiha:2014jxa}
\bibitem{Alishahiha:2014jxa}
M.~Alishahiha, M.~R.~Mohammadi Mozaffar and M.~R.~Tanhayi,
``On the Time Evolution of Holographic n-partite Information,''
JHEP \textbf{09}, 165 (2015)
doi:10.1007/JHEP09(2015)165
[arXiv:1406.7677 [hep-th]].
%47 citations counted in INSPIRE as of 27 Jul 2022
	
	

	
	%\cite{Liu:2019qje}
\bibitem{Liu:2019qje}
P.~Liu, Y.~Ling, C.~Niu and J.~P.~Wu,
``Entanglement of Purification in Holographic Systems,''
JHEP \textbf{09}, 071 (2019)
doi:10.1007/JHEP09(2019)071
[arXiv:1902.02243 [hep-th]].
%41 citations counted in INSPIRE as of 27 Jul 2022
	
	
	

	

	


	
	

\end{thebibliography}
\end{document}